\documentclass[11pt]{article}

\usepackage[]{graphicx,isolatin1}
\usepackage[squaren]{SIunits}
\usepackage{color}

\newcommand{\gfrac}[2]{\displaystyle\frac{#1}{#2}}

\newcommand{\bin} {\textrm{bin}}
\newcommand{\effective} {\textrm{effective}}
\newcommand{\drift} {\textrm{drift}}
\newcommand{\below} {\textrm{below}}
\newcommand{\abo} {\textrm{above}}
\newcommand{\micromegas} {\mu\textrm{M}}
\newcommand{\GEM} {\textrm{GEM}}
\newcommand{\transparency} {\mathcal{T}}
\newcommand{\collection} {\mathcal{C}}
\newcommand{\extraction} {\mathcal{E}}

 \newcommand{\citenum}[1]{\cite{#1}}
\title{HARPO: a TPC as a gamma-ray telescope and polarimeter}

\author{Denis Bernard, 
Philippe Bruel, 
Mickael Frotin, 
Yannick Geerebaert, 
Berrie Giebels, 
\\
Philippe Gros, 
Deirdre Horan, 
Marc Louzir,
Patrick Poilleux, 
Igor Semeniouk, 
Shaobo Wang 
$^{a}$ 
\\
$^{a}$LLR, Ecole Polytechnique, CNRS/IN2P3, 91128 Palaiseau, France
\\
~
\\
Shebli Anvar, 
David Attié, 
Paul Colas, 
Alain Delbart, 
Patrick Sizun
$^{b}$ 
\\
$^{b}$IRFU, CEA Saclay, 91191 Gif-sur-Yvette, France
\\
~
\\
Diego Götz
$^{b,c}$
\\
$^{c}$AIM, CEA/DSM-CNRS-Université Paris Diderot, IRFU/Service d'Astrophysique, 
\\
CEA Saclay, F-91191 Gif-sur-Yvette, France 
}

\begin{document} 

\maketitle

\begin{center}
\large \textbf{ Paper No.	9144-57
\\
Presented at 
\textit{SPIE Astronomical Telescopes + Instrumentation},
\\
\textit{Ultraviolet to gamma ray},
\\
Palais des congrès de Montréal, Montréal, Québec, Canada; 
22 - 27 June 2014 }
\end{center}

\begin{abstract}
A gas Time Projection Chamber can be used for gamma-ray astronomy with
excellent angular-precision and sensitivity to faint sources, and for
polarimetry, through the measurement of photon conversion to $e^+e^-$
pairs. We present the expected performance in simulations and the
recent development of a demonstrator for tests in a polarized photon
beam.
\end{abstract}


\section{INTRODUCTION}
\label{sec:intro} 

Several classes of cosmic sources such as active galactic nuclei
(AGN), pulsars and gamma-ray bursts (GRB) produce huge flows of
gamma rays.
These high-energy photons are produced by non-thermal processes such
as synchrotron radiation and inverse Compton scattering, and provide
insight to understanding the structure of these sources and the
emission mechanisms at work.
In the 10 MeV -- 100 GeV photon energy range, telescopes on space
missions have been using the gamma conversion to $e^+e^-$ pairs in
high-$Z$ converters which are interleaved with particle trackers, to
detect astrophysical sources.

\subsection{Angular resolution}
\label{sub:sec:Angularresolution}

The successive COS-B, EGRET, AGILE and Fermi missions have shown an
impressive improvement in the effective area and in  sensitivity,
but hopes for high sensitivity at low energy 
(e.g.  Ref. \citenum{Schoenfelder}) have been unfulfilled:
even the Fermi-LAT is actually publishing results mainly above 100 MeV.
Fermi is sensitive down to 20 MeV, as is demonstrated by GRB analyses,
for which a loose event selection is possible because of the small
integration time.
But for standard analyses with a long integration time, a much more
stringent selection is required, which affects the effective area
below 1 GeV: 
the effective area is one order of magnitude smaller at 100 MeV than
above 1 GeV
(Fig 14 of Ref. \citenum{Ackermann:2012kna}).
The other key issue at low energy is the strong degradation of the
angular resolution, which makes source finding very difficult,
especially in crowded regions of the sky such as the galactic plane.
The Fermi-LAT differential sensitivity at 100 MeV for class P7SOURCE\_V6 for a
point source for a 3-year exposure, 4 bins per energy decade, 
with a 5~$\sigma$ sensitivity requirement and at least 10 counts per
bin is 
$5 \times 10^{-6} \mega\electronvolt \centi\meter^{-2} \second^{-1}$
for an ``intermediate'' galactic latitude.
%
%
%
This figure, 
$2.4 \times 10^{-6} \mega\electronvolt \centi\meter^{-2} \second^{-1}$, 
 is of the same order of magnitude as that for EGRET
(from Ref.  \citenum{Schoenfelder}), 
after rescaling to the same duration and binning.

The 1 -- 100 MeV sensitivity gap\cite{Schoenfelder} between the
Compton and the pair telescopes has triggered many developments.
On the Compton side, projects aiming at an improvement in the
sensitivity of a factor 10 -- 30 w.r.t. Comptel are in progress 
\cite{Greiner:2011ih,Cube:SPIE2014,AstroMeV}.
This would bring us at the level of 
$\approx 10^{-5} \mega\electronvolt \centi\meter^{-2} \second^{-1}$ at a few MeV.
Bridging this improvement by a detector able to extend at low energy
the excellent high-energy (GeV) sensitivity of the Fermi-LAT telescope
($\approx 10^{-6} \mega\electronvolt \centi\meter^{-2} \second^{-1}$)
is urgently needed.

\subsection{Polarimetry}
\label{sub:sec:Polarimetry}

In contrast with thermal emission and hadronic interactions (which
produce spin-zero $\pi^0$), the emission mechanisms mentioned above
produce a radiation that is linearly polarized to some extent.
Polarization measurements provide an important tool for understanding
the physical processes in astrophysical sources.

{\bf GRBs} are thought to be created either by the explosion of a hypernova
or through the coalescence of two compact objects (e.g.,
neutron stars, white dwarfs, black holes).
In the relativistic jets that are thus produced, synchrotron emission
is thought to be the dominant emission process. Electrons are
accelerated to near light speed by the relativistic shocks and, given
the presence of a strong magnetic field
\cite{Piran:1999:GRB-Fireballs}, the degree of polarization of the
photons emitted by these jets is expected to be very
high. 
Polarization measurements of the emission will therefore provide
information critical to distinguishing between the many emission
models that exist for GRBs (see, for example, Refs. 
\citenum{Waxman:2003:PolGRBs, Granot:2003:PolGRBs,Eichler:2003:PolGRBs,
Lyutikov:2003:PolGRBs}).

Another source class whose study should benefit from polarimetry
studies is {\bf AGN} and, in particular, the {\bf blazar}
subclass, which emits strongly in the MeV - GeV energy range.
These systems have powerful jets with strong magnetic fields that are
thought to contain a large portion of the energy reservoir of the jet
and to play a key role in the variability that is characteristic of
these objects. Blazars have two broad emission components; the first,
at lower energies, is believed to be from synchrotron emission by
electrons and positrons in the strong magnetic field of the jet. The
origin of the second emission component at higher energies has been
the subject of much theoretical modeling.
The proposed models fall into two broad categories, namely, hadronic
and leptonic models. In the leptonic models, the high-energy electrons
and positrons that generate the synchrotron emission, also upscatter
photons to higher energies via the inverse-Compton mechanism,. 
Recently Zhang {\em et al.} have extended their predictions to the
polarization of X and gamma-ray emission of blazars to polarization 
\cite{Zhang:2013bna}.
Using the example of the blazar RX J0648.7+1516, they show that for
the subclass of high-synchrotron-peaked (HSP) blazars, the fraction of
linear polarization $P$ in the X-ray band (2 -- 10 keV) is high and
similar in magnitude for leptonic and hadronic models, while in the $\gamma$ band
(30 -- 200 MeV) it is predicted to be of the order of 70\% for protons
and to be completely washed out for electrons (their Fig. 5).

Gamma-ray polarimetry also turns out to be the most sensitive tool to
test possible {\bf Lorentz invariance violation} (LIV), considered in
building a quantized theory of gravitation.
Within the framework of an effective field theory, a birefringence
effect of the vacuum is predicted \cite{LIV}.
The linear polarization direction would be rotated through an
energy-dependent angle, due to different phase velocities for opposite
helicities.
This vacuum birefringence would rotate the polarization direction of
monochromatic radiation, or could depolarize linearly polarized
radiation composed of a spread of energies.
Due to the quadratic dependence of the deviation angle on the photon
energy, $\theta \propto E^2$, extending the polarization measurements
from the presently available range covered by Compton polarimeters
(0.1 - 1.0 MeV) to higher energies would lead to an improved
sensitivity to LIV.

\section{HARPO: a TPC as a gamma-ray telescope and polarimeter}
\label{sec:HARPO} 

The HARPO (Hermetic ARgon POlarimeter) project aims at characterizing
the time projection chamber (TPC) technology as a high angular resolution
and polarimeter telescope in the MeV - GeV energy range.

\subsection{Angular resolution}
\label{sub:sec:HARPO:Angularresolution}

The reconstruction of the incident direction of a photon that converts
in a detector is affected by a series of effects that contribute to
the angular resolution.

\begin{itemize}
\item {\bf Multiple scattering.}
The main factor affecting 
 the lowest part of the energy
range considered here is  multiple scattering of the conversion
electrons. 
Based on earlier works by Gluckstern \cite{Gluckstern} and Innes
\cite{Innes:1992ge} on the determination of track parameters with
optimal fits in the presence of multiple scattering, we have obtained\cite{Bernard:2012uf}
a parametrization of this (RMS) contribution as:

\begin{equation}
\sigma_{\theta} 
 \approx 
(2 \sigma)^{1/4} l^{1/8} X_0^{-3/8} (E/p_0)^{-3/4}.
 \label{eq:Vbb:p:2}
\end{equation}

where $E$ is the photon energy, 
$p_0 = 13.6 \, \mega\electronvolt/c$ is the usual multiple scattering
parameter,
$X_0$ is the radiation length of the TPC material and 
$\sigma$ and $l$ are the single point space resolution and the
sampling pitch of the TPC, respectively.
In deriving this expression we have neglected the small logarithmic
correction factor that occurs in the expression for the multiple
scattering angle
 (eq. (30.15) of Ref. \citenum{Beringer:1900zz}).
Also an additional factor, close to unity, that arises in the
computation of this contribution to the photon resolution, from the
expression for each of the two tracks, is neglected
 (Fig 1 of Ref. \citenum{Bernard:2012uf}).
We have validated \cite{Bernard:2013jea} the effectiveness of such
optimal fits fitting the track with a Kalman filter, using the
prescriptions of Ref.\citenum{Fruhwirth:1987fm}, see Fig. \ref{fig:2} left.
The $E^{-3/4}$ dependance of $\sigma_{\theta}$
(eq. (\ref{eq:Vbb:p:2})) can be compared to what has been observed
experimentally, for example $E^{-0.78}$ for the Fermi-LAT
\cite{Ackermann:2012kna}.

\item {\bf Nucleus recoil.} 
In the case of  photon conversion in the
 field of a nucleus, the path length of the recoiling nucleus in the
 detector is extremely short and its direction can't be measured.
That missing contribution to the photon momentum induces a
contribution to the angular resolution.
Our study \cite{Bernard:2012uf} shows that while the most probable
angular shift varies like $E^{-2}$
and becomes negligible above a few MeV, the increase of the fraction
of  events in the high recoil tail above the most probable value
gives a different dependance to the the 68\% containment angle, 
$\theta_{68}$, which is the relevant figure in astronomy.
We have discovered\cite{Bernard:2012uf} that $\theta_{68}$ varies like
$E^{-5/4}$.
\end{itemize}

\begin{figure} [th]
 \begin{center} 
 \includegraphics[width=0.50\linewidth]{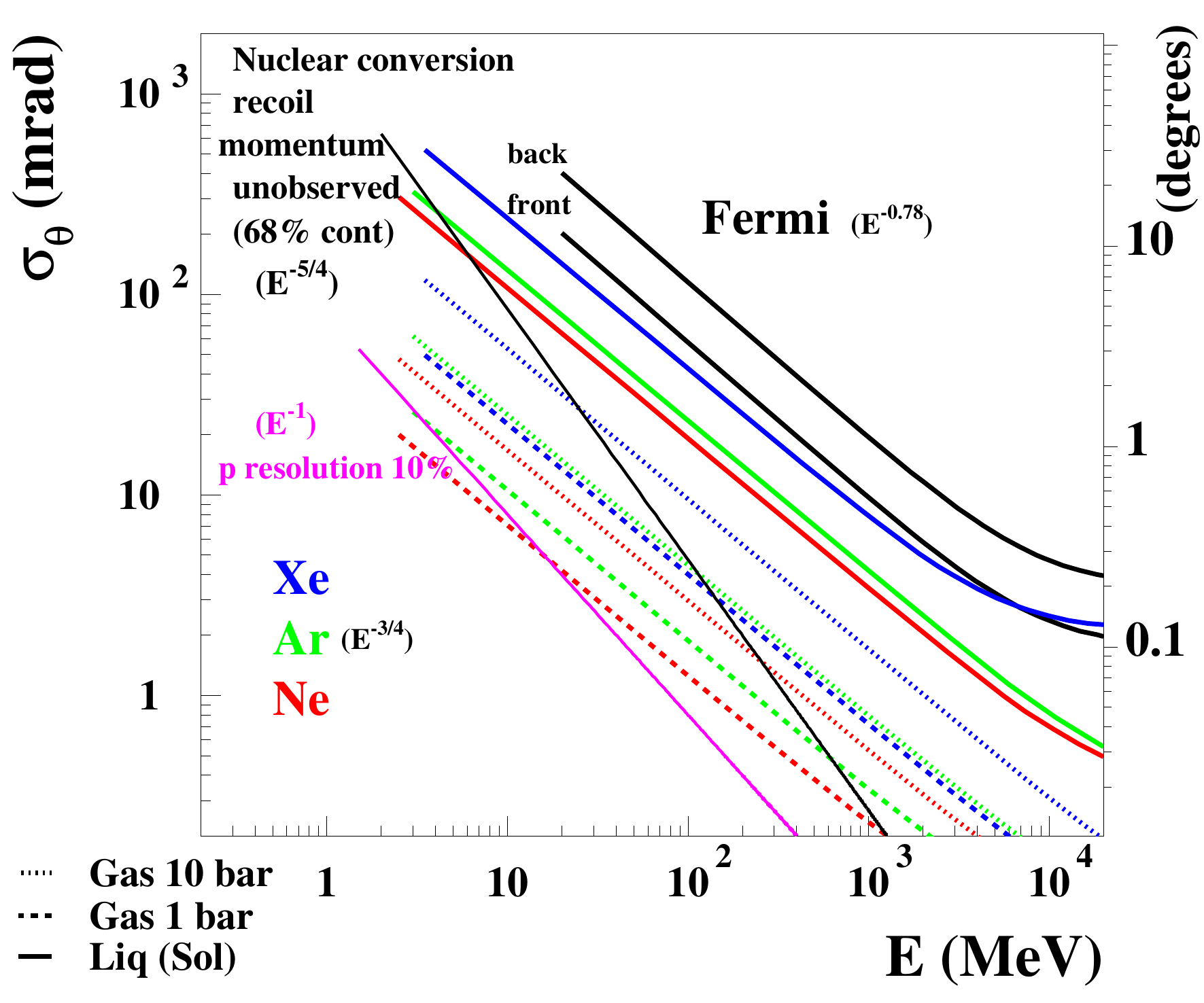}
\hfill
\includegraphics[width=0.43\linewidth]{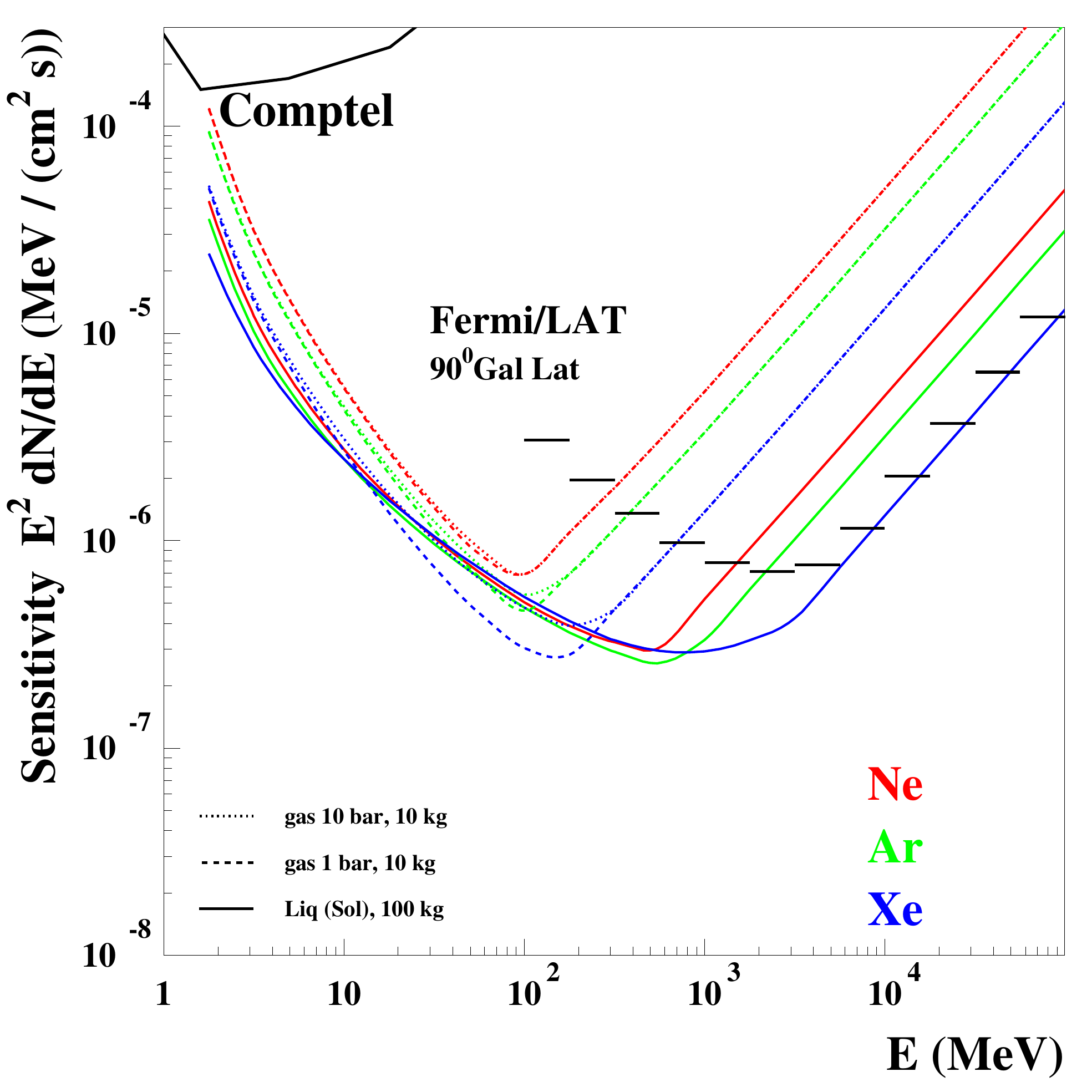}
\caption{\label{fig:1}
Left: Various contributions to the photon angular resolution.
\cite{Bernard:2012uf}.
Right: Variation of the differential sensitivity as a function of
energy \cite{Bernard:2012uf} compared to the $90^\circ$ galactic
latitude performance of the Fermi-LAT \cite{Ackermann:2012kna} and of
the Compton telescope COMPTEL \cite{Schoenfelder}.
}
\end{center}
\end{figure}

As a consequence of the stronger $E$ dependance, the kinematic limit
dominates just above threshold, while multiple scattering dominates at
intermediate energies.
This is shown on Fig. \ref{fig:1} left.
The plot shows the angular resolution for 
a point resolution $\sigma = 0.1~\milli\meter$
and a sampling $l = 1.0~\milli\meter$, for various TPC material 
(neon, argon, xenon) and densities (1 or 10 bar gas, 
liquid (Ar, Xe) or solid (Ne)).
We see that in most of the energy range, an improvement of the angular
resolution by more than one order of magnitude is within reach with
respect to the Fermi-LAT 
(thicker black lines for ``front'' and ``back'' events
\cite{Ackermann:2012kna}).

\subsection{Conversion efficiency, effective area}
\label{sub:sec:Conversionefficiency-effectivearea}

In contrast with a ``thick'' detector, in which the probability of
photon conversion is close to unity, in the present case of a ``thin''
detector, the efficiency is much smaller.
But the relevant criterion is the effective area which is here
proportional to the sensitive mass and to the photon attenuation 
and that is not that small
(larger than $1~\meter^2 / \ton$ above 10 MeV for argon).
A gaseous detector in orbit will certainly contain less than a ton of
sensitive material
($17~\kilo\gram/\meter^3$ for argon at 10 bar), 
but the point-like source sensitivity turns out to be better than that
of the Fermi-LAT in the range 3 -- 300 MeV for a gas detector 
(Fig. \ref{fig:1} right and Ref. \citenum{Bernard:2012uf}).
We see that a 10 kg gas TPC and a 100 kg liquid TPC have similar
sensitivities in that energy range, and that the choice of the target
meterial is not critical, as the larger effective area for a large $Z$
and the better angular resolution for low $Z$ cancel almost exactly in
the expression for the sensitivity.

\subsection{Momentum measurement}
\label{sub:sec:Momentummeasurement}

The measurement of the energy of an incoming photon needs information on
 the momentum $p$ of each electron track.
To do so a transition radiation detector (TRD) 
or a magnetic spectrometer 
or even worse, as far as the mission mass budget is concerned, an
electromagnetic calorimeter has to be used.

\begin{figure}[th]
\begin{center}
 \includegraphics[width=0.6\linewidth]{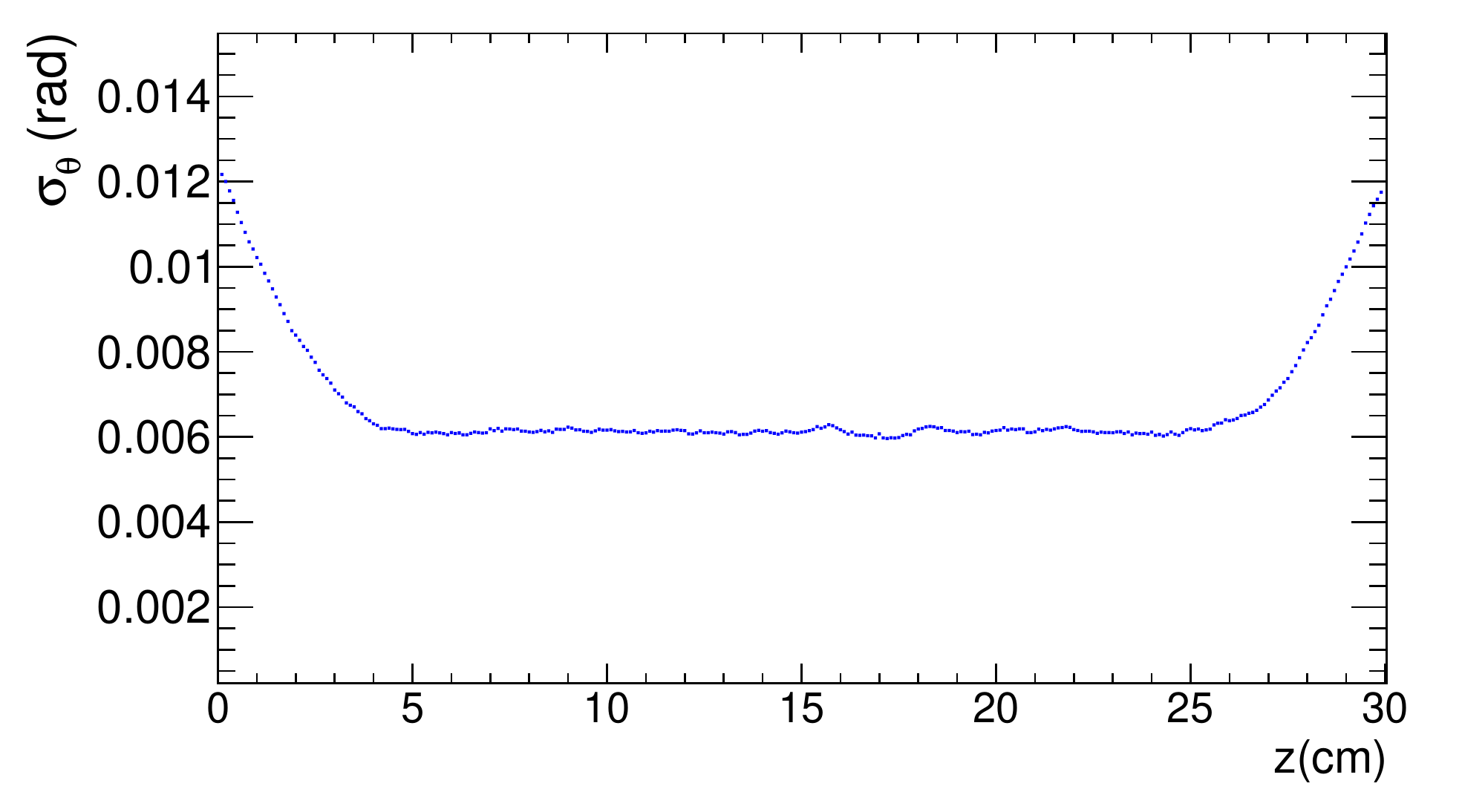}
\hfill
\includegraphics[width=0.38\linewidth]{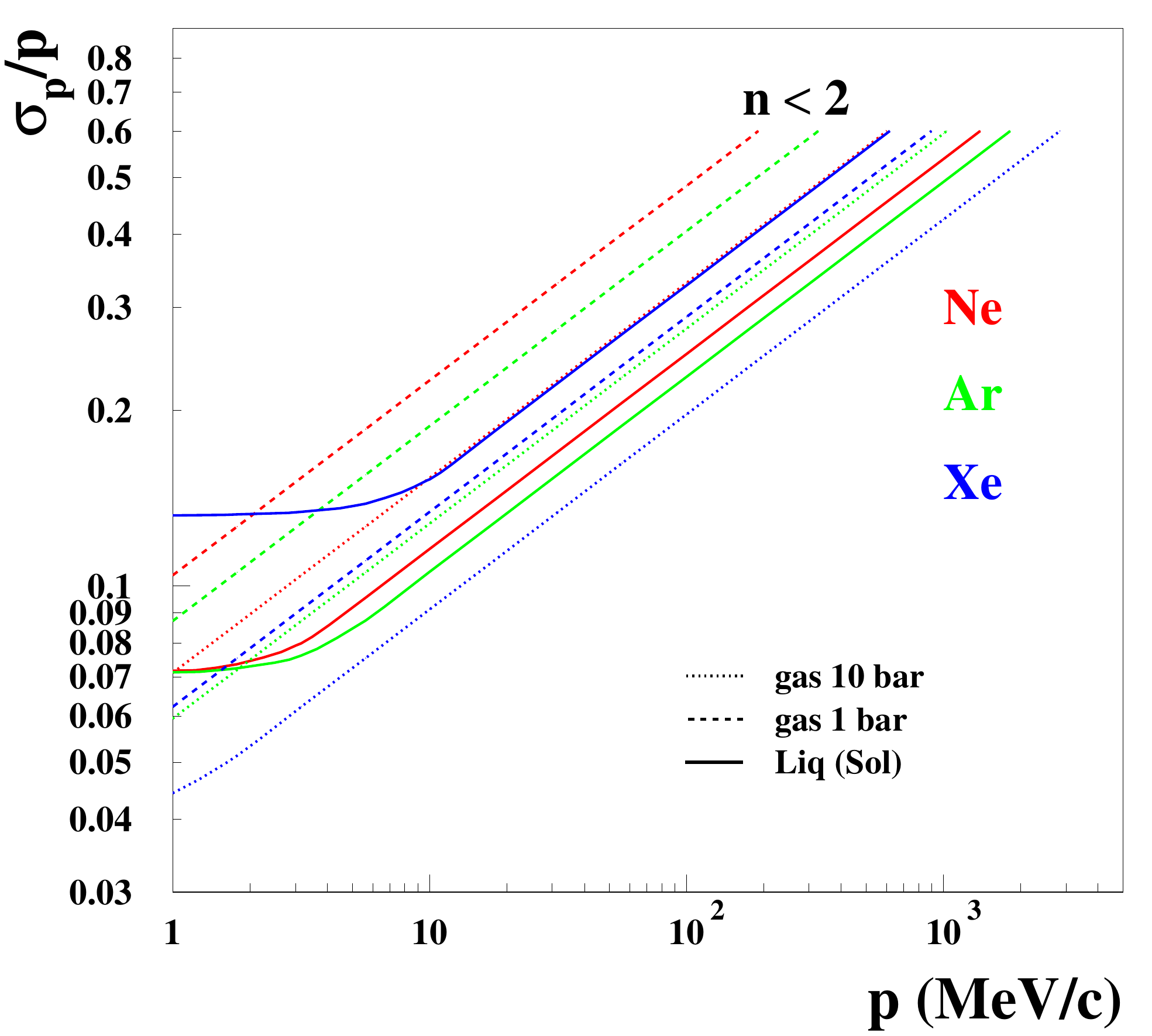}
\caption{Left: 
Single track angle RMS uncertainty of a Kalman filter fit
 from a sample of 5000 simulated $40\,\mega\electronvolt/c$ tracks in
 5\,bar argon with $\sigma = l = 0.1~\centi\meter$;
Eqs. (\ref{eq:Vbb:p:2}) predicts an RMS of $12.2\,\milli\radian$
 \cite{Bernard:2013jea}.
Right:
Relative track momentum resolution for optimal sampling as 
a function of track momentum for a short TPC (30 cm gas, 10 cm liquid/solid)
\cite{Bernard:2012uf}.
\label{fig:2}
}
\end{center}
\end{figure}

But the TPC itself has some measurement potential, as has been used
repeatedly (for example in emulsions) since the pioneering works of
d'Espagnat \cite{dEspagnat}
and Molière \cite{Moliere}:
Given the $1/p$ dependance of the RMS multiple scattering angle, the
multiple measurement of angular deflections in the detector provides a
measurement of the track momentum.

Here again, the resolution involves contributions from multiple
scattering and from the detector resolution. 
Optimizing the ``cell'' length on which the measurement is performed,
that is, regrouping the TPC sampling in an optimal way, and neglecting
conservatively the improvement of the spatial resolution to be
expected from the regrouping of several sampling in one cell, we
obtain \cite{Bernard:2012uf} the results presented in Fig. \ref{fig:2} right.
As expected the method is useable at low energy ($\sigma_p /p > 0.3$
for $p > 100~\mega\electronvolt/c$, over 30 cm in 10 bar argon).

\subsection{Polarimetry}
\label{sub:sec:HARPO:Polarimetry}

Polarimetry, the measurement of the fraction of linear polarization of
cosmic sources, is performed for hard X-rays and soft gamma-rays
using Compton telescopes, but the sensitivity to polarization is
extremely low above 1 MeV
\cite{McConnell-Ryan},
due to the asymptotic (at high energy) $m/E$ dependance of the
polarization asymmetry
 \cite{Bernard:2012uf}.

\subsubsection{Slabs}
\label{sub:sub:sec:slabs}

Hopes to use pair conversion face the hurdle of measuring the
azimuthal angle of the conversion plane before multiple scattering
ruins this angular information
\cite{Kotov,Mattox,Bloser:2003gb,Hunter:2013wla}.
In a detector formed as a multi-slab series of converter / tracker
combinations, multiple scattering in the conversion slab induces an
angular deflection of the electron tracks that blurs the azimuthal
angle $\phi$ reconstructed from them.
Approximating the two-track opening angle to be equal to the most
probable value, produces a dilution\footnote{Note 
that this expression for the dilution is energy
 independent: At high energy, the electrons suffer less multiple
 scattering but the pair's opening angle decreases as $m/E$.}
$D$ of the polarization asymmetry $D = e^{-2 \sigma_{\phi}^2}$ with
$\sigma_{\phi} \approx 14 \sqrt{x/X_0}$, where $x$ is the track pathlength
in the slab. \cite{Kotov,Mattox}.
This would lead to the use of slabs with an unrealistically small
thickness, that is to an unreasonably large number of tracker
layers.

To overcome this difficulty, the use of triplet\cite{Perrin1933}
conversion has been considered: in the case of the
conversion of a photon in the electric field of an electron of the
detector, the path length of the recoiling electron can be visible in
the detector, despite the small recoil energy, thanks to the smallness
of the electron mass relative to that of an ion.
The azimuthal direction of the recoil carries approximately the same
information as the azimuthal angle of the pair plane for polarimetry,
but at the cost of the detection and of the reconstruction of very low
momentum electrons (check Fig. 3 of Ref. \citenum{Bernard:2012uf}.).
This has triggered many studies
(among which Refs. 
\citenum{SuhBethe1959,Endo:1992nq,Boldyshev:1994bs,Iparraguirre:2011zz,Depaola:2009zz,Iparraguirre:2014lia}
(and references therein))
since Votruba first computed the differential cross section
\cite{Votruba1948}.
But as we shall see \cite{Bernard:2013jea}, the sensitivity to
polarimetry of triplet conversion using practical detectors for actual
cosmic sources is rather low.

We have studied in detail the parameters that determine the potential
of a TPC as a pair-conversion polarimeter in
Ref. \citenum{Bernard:2013jea}, considering both triplet and
``nuclear'' conversions, and focusing interest on the MeV -- GeV
energy range which, in terms of pair-conversion, is the low energy
range, but where most of the signal from cosmic sources is expected.

\begin{figure}[th]
\begin{center}
 \includegraphics[width=0.7\linewidth]{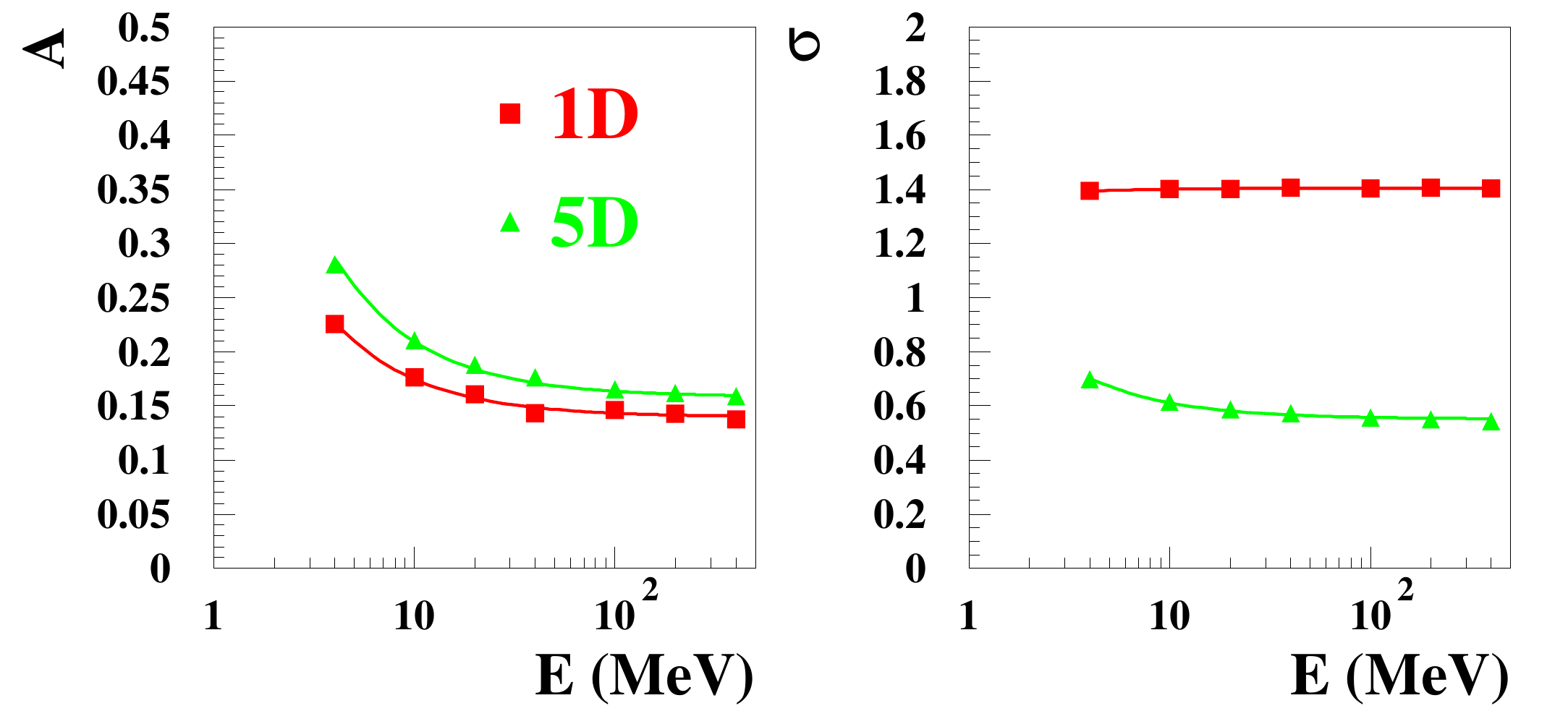}
\caption{Average (left) and RMS (right) values of the
 polarization asymmetry without dilution, estimated with a 1D and 5 D
 optimal variable \cite{Bernard:2013jea}.
\label{fig:1d5d:fit}
}
\end{center}
\end{figure}

\subsubsection{A full, exact, polarized event generator }
\label{sub:sub:sec:EvtGen}

Most of the photon conversion event generators on the 
market \cite{EGSnrc287,Depaola:2000qd,Depaola:2009zz} don't sample the
five-dimensional differential cross section, but instead use a
factorized approximation of a 1D or 2D projection of it.
Also low-energy approximations are often used.
A number of ``polarized'' event generators only take polarization into
account for lower energy processes such as Compton scattering and the
photo-electric effect \cite{Cirrone:2010zz}.

We have interfaced the HELAS amplitude calculator
\cite{Murayama:1992gi}
with the SPRING event generator \cite{Kawabata:1995th}
to create a full (5-dimensional), exact (down to threshold), polarized
event generator, that includes all diagrams either for ``nuclear'' or
for ``triplet'' conversion.
The validation of this event generator w.r.t. 1D
distributions published in the past is described to some extent in
section 3 of Ref. \citenum{Bernard:2013jea}.
(Check, for example, the photon energy variation of the triplet
cross section above a recoil momentum threshold (Fig .6 of
Ref. \citenum{Bernard:2013jea}) compared to the high-energy asymptotic
expression of Ref. \citenum{Iparraguirre:2011zz}).
We have also checked the correctness of the amplitude calculation
itself by comparison with the Bethe-Heitler (BH) differential
cross section \cite{Heitler1954}.

We have explored the actual potential of additional kinematic cuts
devised to enlarge the effective value of the polarization asymmetry,
an idea that has lead to intense efforts in the past (e.g. 
 \cite{Endo:1992nq,Asai-Skopik,Adamyan:2006im}). 
When the loss of statistics induced by the selection is taken into
account, the net benefit of applying such a selection, if any, stays
minimal.
In contrast with polarimeters on man-made accelerator photon beam
lines, for which statistics are not an issue
\cite{deJager:2007nf,Wojtsekhowski:2003gv}, for operation in space a
high selection efficiency is mandatory.

We have explored the potential of the use of optimal variables, to best 
extract the information on the polarization contained in the 5D
differential cross section.
Using a 1D optimal variable doesn't improve the precision of the
measurement of $P$ much, compared to a simple fit of the distribution
of the azimuthal angle.
In the case of the full 5D optimal variable, the gain in precision is
of a factor of about 2
(Fig. \ref{fig:1d5d:fit})
 at the cost of an increased complexity and of an increased
 sensitivity to resolution effects.

\subsubsection{Polarimetry in a real detector: multiple scattering and optimal fits}
\label{sub:sub:sec:Real:Det:Polarimetry}

We have first studied the traditional ``multi-slab'' detector geometry
with our full event generator, obtaining a dependance of the dilution
as a function of the normalized thickness $t \equiv x / X_0$
completely different from the most-probable opening angle value
mentioned above \footnote{
Note, though, that intermediate values are similar for the two
calculations ($D\approx 1/2$ for $t \approx 10^{-3}$).}.

\begin{figure}[th]
\begin{center}
 \includegraphics[width=0.435\linewidth]{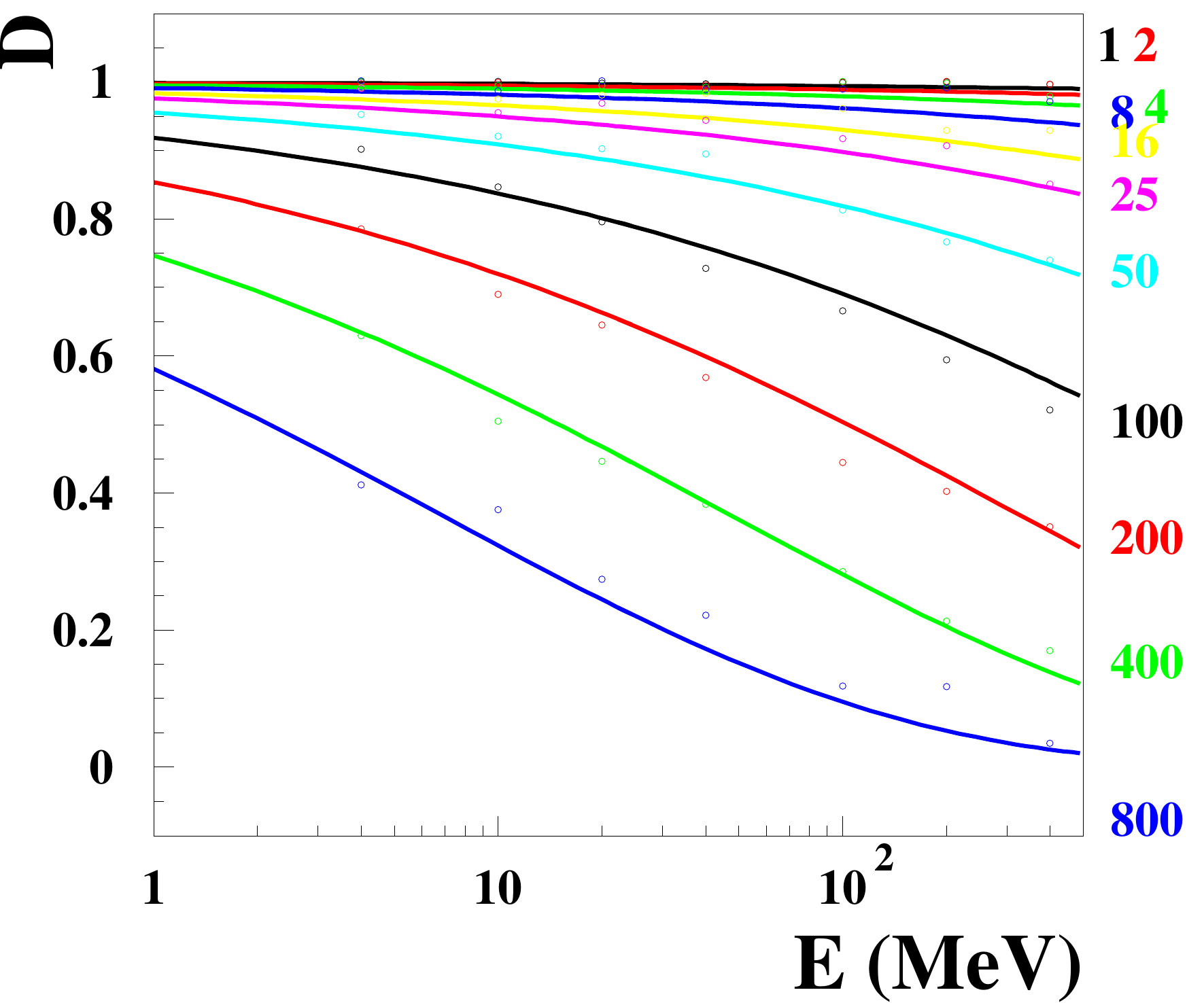}
\hfill
 \includegraphics[width=0.48\linewidth]{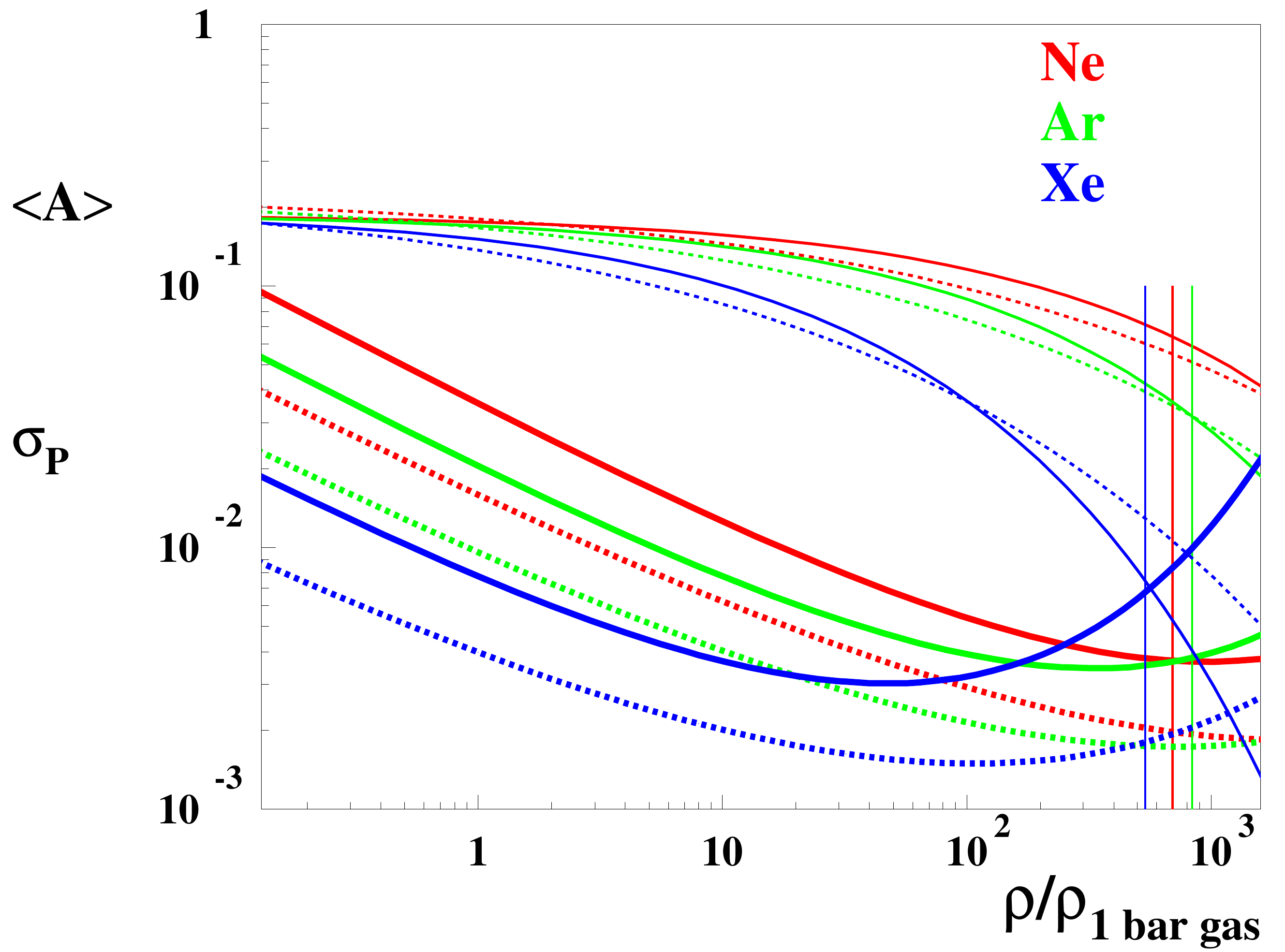}
\caption{
Left:
Energy variation of the dilution $D$ of the measurement of
 $P$, for various values of the parameter $p_1$ that parametrize the
 single-track angular resolution in a thin detector (the values of
 $p_1$ (keV) are listed on the right of each plot
(1D optimal variable) \cite{Bernard:2013jea}.
Right:
Average polarization asymmetry (thin line) and polarization fraction
precision (thick line) as a function of detector density normalized
to the 1\,bar gas density, for a $1\,\meter^3$ sensitive volume detector
exposed for 1\,year and a Crab-like source \cite{Bernard:2013jea}
(nuclear conversion, $\eta = \epsilon = 1$).
1D (solid line) and 5D (dashed line) weight.
The vertical lines show the density of the liquid phase.
\label{fig:4}
}
\end{center}
\end{figure}

We then extend the study to optimal fits. To express the results, we
found it interesting to recast
eq. (\ref{eq:Vbb:p:2}) into 
$ \sigma_{\theta} = (p/p_1)^{-3/4}$
with:
$ p_1 = p_0 \left(\gfrac{4 \sigma^2 l}{X_0^3} \right)^{1/6}$, 
therefore characterizing a detector, with regard to multiple
scattering, by the single parameter $p_1$, the ``characteristic multiple
scattering momentum'' of that detector.
For example we obtain $p_1 = 50\,\kilo\electronvolt/c$ for 1\,bar argon,
and
$p_1 = 1.45\,\mega\electronvolt/c$ for liquid argon
(with here $\sigma = l = 0.1~\centi\meter$).
Due to the $p^{-3/4}$ dependance of the single track angular
resolution for optimal fits, the dilution is not a constant anymore
and gets worse at high energy (Fig. \ref{fig:4} left).

The precision on a Crab-like source, obtained with a $1~\meter^3$ TPC
filled with 5 bar of argon-based gas, with a one full year exposure
and 100 \% efficiency, would be of 1.0 \% (Fig. \ref{fig:4} right).
Taking into account minimal kinematical cuts, on the opening angle, on
the momentum of each of the two leptons, and on the angular distance
to the emitting source, it would increase to 1.4 \%.

\begin{figure}[th]
\begin{center}
 \includegraphics[width=0.6\linewidth]{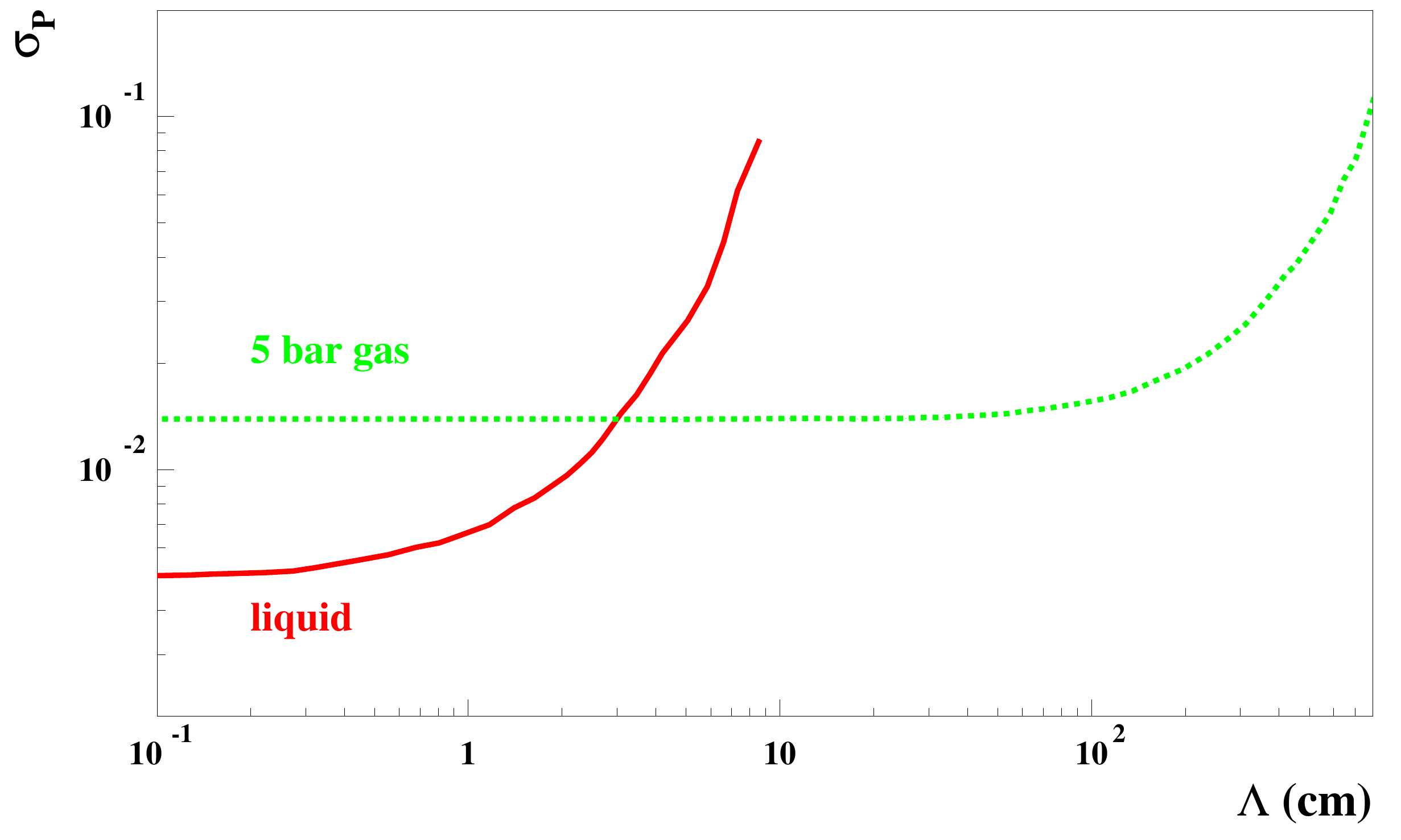}
\caption{ Comparison of the precision $\sigma_P$ of the measurement of the polarization of a Crab-like source, for a 5 bar gas and a liquid TPC, as a function of the track length limit (1 year, 1$\meter^3$, exposure fraction $\eta = 1$).
\label{fig:gas:liq}}
\end{center}
\end{figure}

\begin{figure}[th]
\begin{center}
 \includegraphics[width=0.8\linewidth]{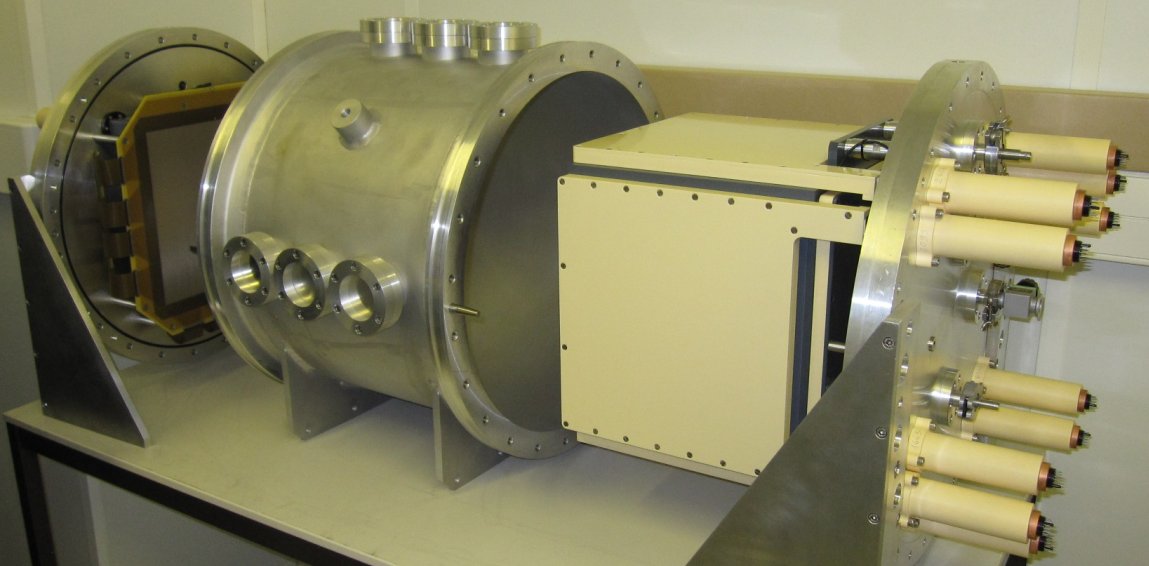}
\caption{
The demonstrator being mounted (2011). From left to right :  
micromegas, pressure vessel, TPC surrounded by trigger scintillators, PMTs. 
\label{fig:det}}
\end{center}
\end{figure}

It is  clear from Fig. \ref{fig:4} left that performing polarimetry
with pair conversions in a dense detector such as a liquid argon TPC
\cite{Caliandro:2013kba}, a solid neon or silicon TPC, 
a stack of silicon detectors \cite{Morselli:2014fua}
or a
scintillator cube \cite{Cube:SPIE2014} will be extremely difficult,
the high value of $p_1$ making only the very lowest part of the energy
spectrum (barely) sensitive to the polarization of the incoming
radiation, precisely in the enegy range where the angular resolution
of these dense detectors is disastrous (almost as bad as that of the
Fermi-LAT, see Fig. \ref{fig:1}), and the triggerability
of the telescope and photon selection from a given source, very
problematic.
In practice the detector parameter that determines the polarimetric
performance of a dense detector is the minimum length $\Lambda$ for
which tracks can be reconstructed: for a liquid-argon TPC to
out-perform a 5 bar argon gas TPC would need event detection / trigger /
reconstruction with tracks as short as a centimeter
(Fig. \ref{fig:gas:liq}).

The schema of a telescope intended for a space mission is shown in
Fig. \ref{fig:schema:flight:detector}.
It consists of a $z$ pile of 3 layers, each layer of 2 back-to-back
modules\footnote{The
back-to-back positioning of two TPC modules, with a
 thin common metallic grid as a cathode as used for Aleph 
 \cite{Decamp:1990jra} for example, would here make the handling of
 the drift high-voltage easier.}, with
each module having a collection plane segmented into 
$6 \times 6$ $(x,y)=(33~\centi\meter)^2$, 
$z=50~\centi\meter$ blocks. 
The sensitive volume is $12~\meter^3$ and for a realistic estimate of
$\eta \times \epsilon = 0.1$, the sensitivity is close to  1 \%.
With $2.6~\milli\meter$ wide strips at a $5.2~\milli\meter$ pitch that
is 64 channels (one chip) for each direction $(x,y)$,
this $\approx$ one ton telescope would need 432 such
chips\footnote{The power consumption of that chip, in the present
  version which is not optimized for space application, is 1~W.}.

\section{HARPO: the experimental project}
\label{sec:HARPO:exp} 

\subsection{The ``ground'' demonstrator}

A ``ground'' demonstrator was designed in 2010, 
built in 2011 \cite{Bernard:2012tda,Bernard:2012em}
and tested with cosmic rays in 2012 \cite{Bernard:2012jy}.
The detector is a $(30~\centi\meter)^3$ cubic TPC surrounded with 6
scintillator plates that provide an external trigger.
The TPC endplate includes a bulk \cite{Giomataris:2004aa}
 micromegas \cite{Giomataris:1995fq}
 amplification system: a metallic mesh is held at $128~\micro\meter$ above 
a segmented collection plane made of a PCB covered over an instrumented
width of $288~\milli\meter$ by two crossed series of $35~\micro\meter$ thick copper strips, each at a
pitch of $1~\milli\meter$.
When the mesh is put at a high voltage w.r.t. the collection
plane, amplification takes place in the thin  space between them.
The signals are extracted from the pressure vessel and sampled by the
electronics at a rate of $1/(n \times 10 ~\nano\second)$, $n>1$, based on
the AFTER chip, that was originally designed for the T2K experiment
\cite{Baron:2008zza,T2K_ELEC,DenisCalvet}.
It has  a DAQ rate limited to $< 180~\hertz$.

Cosmic-ray tests performed with a ``T2K'' gas mixture (Ar:95
Isobutane:2 CF4:3 \%), mainly at a pressure of 2 bar have shown that this
demonstrator has excellent tracking properties \cite{Bernard:2012jy}.
After zero-suppression was (optionally) performed in the electronics, 
data analysis included: 
\begin{itemize}
\item pixel thresholding; 

\item pixel clustering; 

\item track pattern-recognition using combinatorial Hough transform;

\item the matching of the $(x, t)$ and $(y,t)$ tracks, based on the
 close similarity of the time sampling of the same track signal by
 the two reading systems (along directions $x$ and $y$);

\item track fitting.
\end{itemize}

\subsection{The 2013 upgrade}

The 2012 tests have shown that with the $0.4~\milli\meter$ narrow
collecting strips that we are using, the micromegas alone does not
provide sufficient amplification for routine operation in a safe
configuration.
We have therefore complemented the micromegas with two layers of Gas
Electron Multiplier (GEM) \cite{Sauli:1997qp}.
GEM consist of a $50~\micro\meter$ thick kapton foil with a copper
layer on each face and pierced with holes of $70~\micro\meter$
diameter, with a $140~\micro\meter$ pitch.
When a high voltage is applied between the two faces, amplification
takes place inside the holes.

\subsubsection{Test of $\micromegas + \GEM$ combinations with a $^{55}$Fe radioactive source}

We characterized the combination of a micromegas and of either one or
two GEM in a (Ar:95 Isobutane:5 \%) gas mixture at atmospheric
pressure.
This is done in a dedicated test setup, using a $^{55}$Fe
source\cite{PhGros:Feb2014:RD51}. 
The successive amplification steps are kept at a distance of 
$2~\milli\meter$ from each other by spacers.

\begin{figure} [th]
\begin{center}
\hfill
 \includegraphics[width=0.28\linewidth]{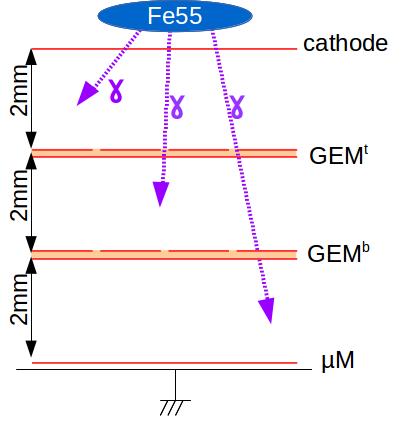}
\hfill
 \includegraphics[width=0.46\linewidth]{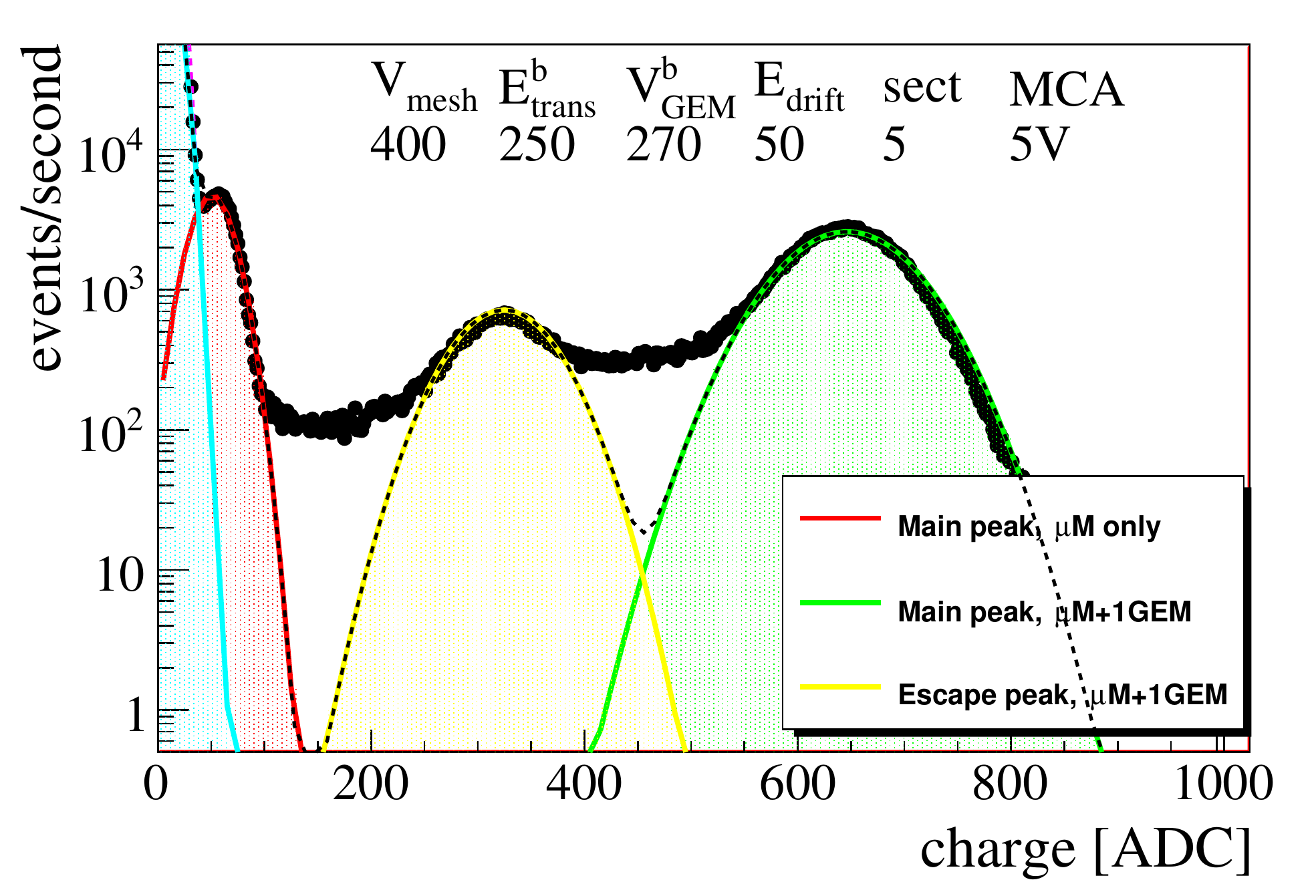}
\hfill
~
\caption{Left: schema of the test set-up.
Right:
Measured spectrum from a $^{55}$Fe source for a 
$\micromegas + 1 ~ \GEM$ combination. 
Three peaks are visible: two ionization peak of $^{55}$Fe (main and
escape) amplified through one GEM and one micromegas, and the main
ionization peak with micromegas amplification only\cite{PhGros:Feb2014:RD51}. 
\label{fig:spectrum}}
\end{center}
\end{figure}

\begin{figure} [th]
\begin{center}
 \includegraphics[width=0.48\linewidth]{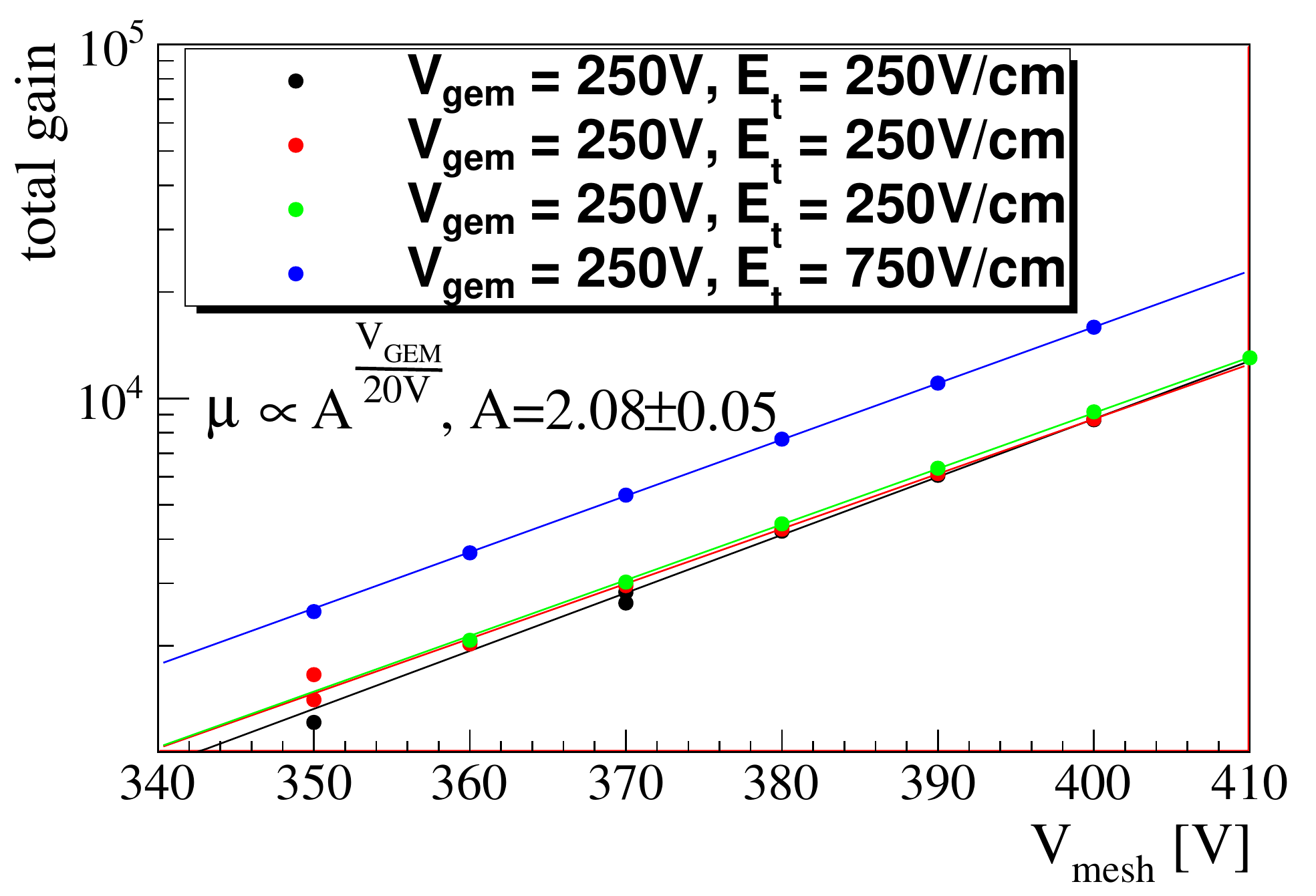}
\hfill
 \includegraphics[width=0.48\linewidth]{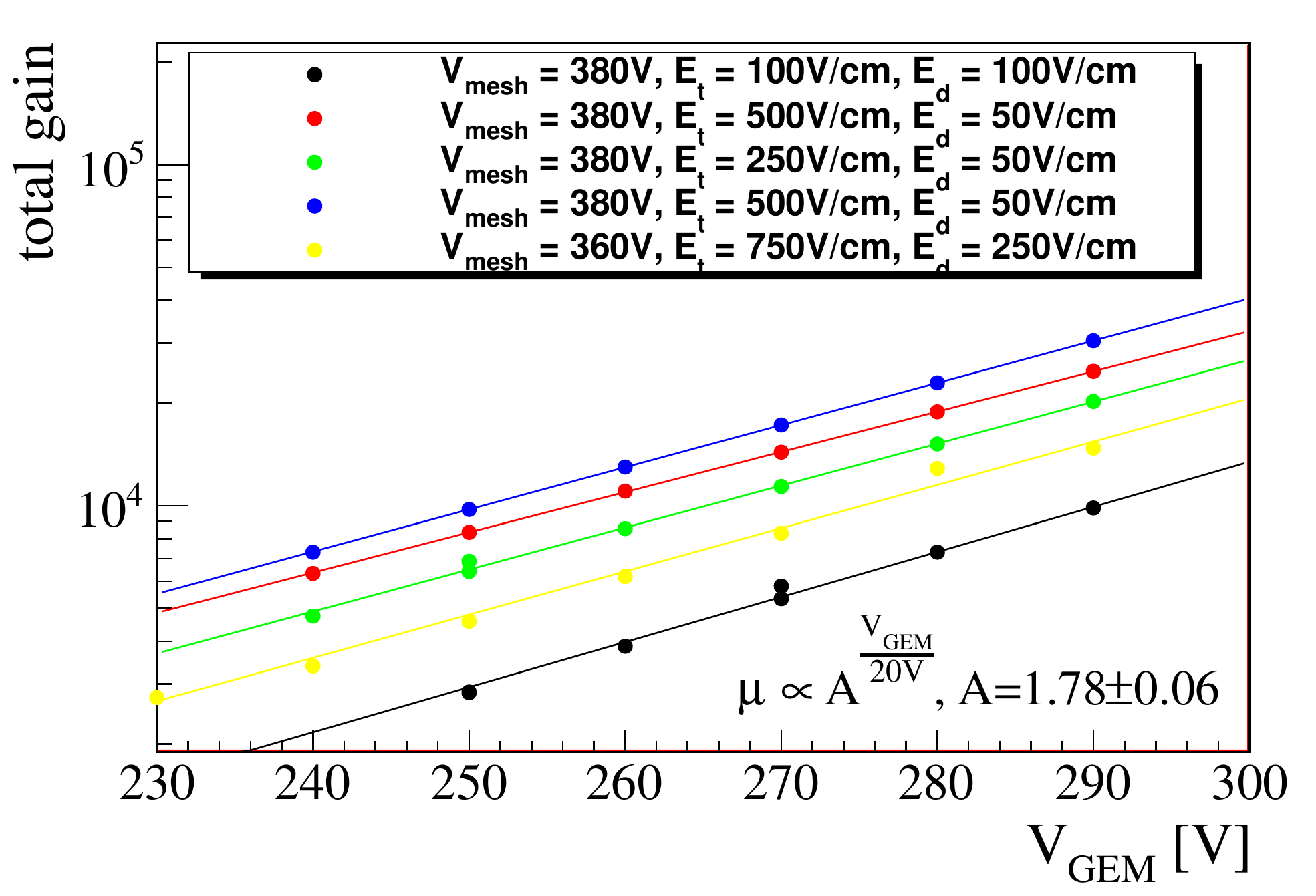}
\caption{Characterization of a $\micromegas + 2 ~ \GEM$ combination in
 a 1 bar (Ar:95 Isobutane 5 \%) gas mixture and X-rays from a $^{55}$Fe
 radioactive source\cite{PhGros:Feb2014:RD51}. }
\label{fig:VScans}
\end{center}
\end{figure}

In argon, the X-rays from a $^{55}$Fe source deposit $5.9$~keV (main
peak) or $2.7$~keV (escape peak) in ionization.
This conversion can happen either above or below a given GEM sheet.
The ionization electrons are therefore amplified by that GEM or
not.
A typical measured spectrum is shown in Fig.~\ref{fig:spectrum} right, where
we can see the main and escape peaks with amplification from the
micromegas and one GEM, and the main peak with micromegas
amplification only.
The ratio of the two main peaks provides a precise measurement of the
absolute GEM amplification gain.
These measurements are used to test the dependence of the gain with
the values of the electric fields in the setup.
Figure~\ref{fig:VScans} shows that the amplification gain grows
exponentially with the voltage applied for both the micromegas and the GEM.
Moreover, the slope is independent of the other parameters.

The effective gain of the amplification chain depends also on the ``transfer''
electric field on each side of the GEM and above the micromegas.
This is usually described as transparency
($\transparency_{\micromegas}$) for micromegas, and collection
($\collection_{\GEM}$) and extraction ($\extraction_{\GEM}$)
efficiency for GEM.
At a fixed amplification voltage, we can assume that the micromegas
transparency and the GEM collection efficiency depend on the value of the electric field above
the amplification system considered, while the GEM extraction efficiency depends on the value of the 
field below the GEM. The respective gains can then be decomposed as
follows:
\begin{eqnarray}
g^{\effective}_{\micromegas} = g_{\micromegas}(V_{\micromegas}) \times \transparency_{\micromegas}(E_{\abo}) \\
g^{\effective}_{\GEM} = \extraction_{\GEM}(E_{\below}) \times g_{\GEM}(V_{\GEM}) \times \collection_{\GEM}(E_{\abo}) 
\end{eqnarray}

\begin{figure} [th]
\begin{center}
 \includegraphics[width=0.48\linewidth]{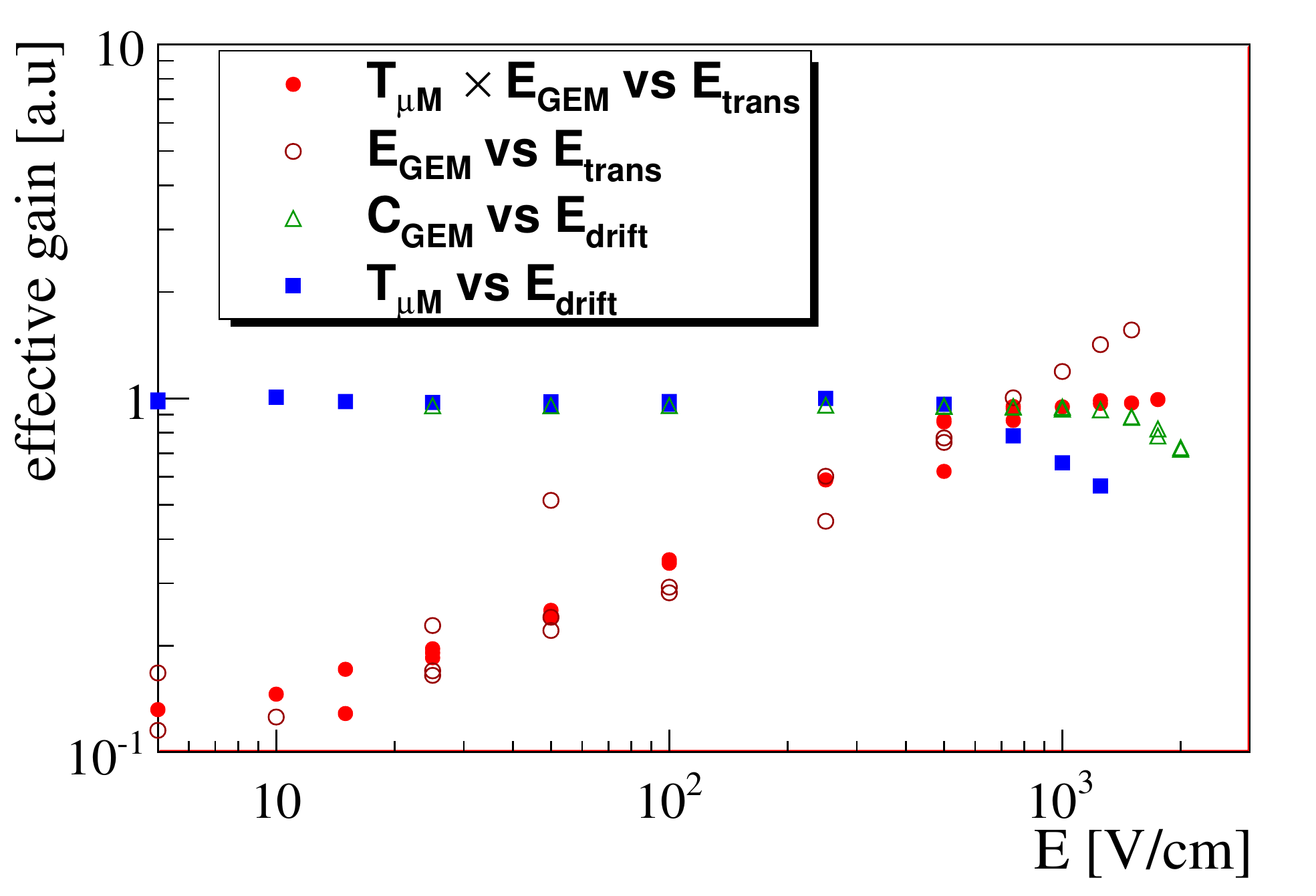}
 \caption{Measurement of the micromegas transparency $\transparency$
 and of the GEM collection $\collection$ and extraction
 $\extraction$ efficiencies as a function of the relevant electric
 field\cite{PhGros:Feb2014:RD51}. The normalisation is arbitrary.}
 \label{fig:EScans}
\end{center}
\end{figure}

By measuring the gain variations with each of the fields, we can
extract these three parameters (Fig~\ref{fig:EScans}).
The micromegas transparency $\transparency$ and GEM collection
$\collection$ show a clear plateau, that probably corresponds to 100\%.
The GEM extraction does not reach a plateau on the field range that we
have explored, and therefore we can't determine the absolute scale of
the GEM transparency.

\subsubsection{Cosmic ray tests of the up-graded demonstrator}

\begin{figure} [th]
\begin{center}
 \includegraphics[width=0.32\linewidth]{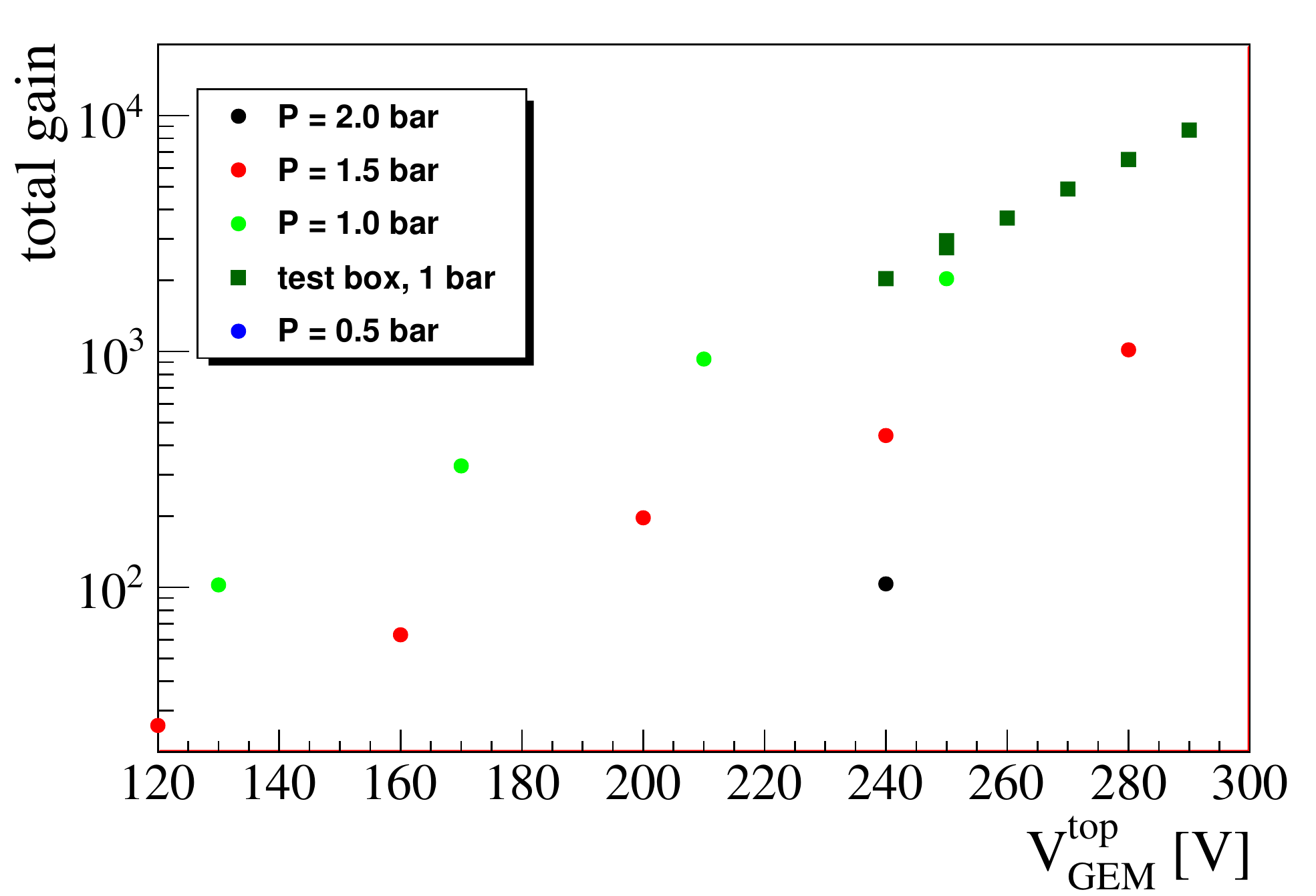}
 \includegraphics[width=0.32\linewidth]{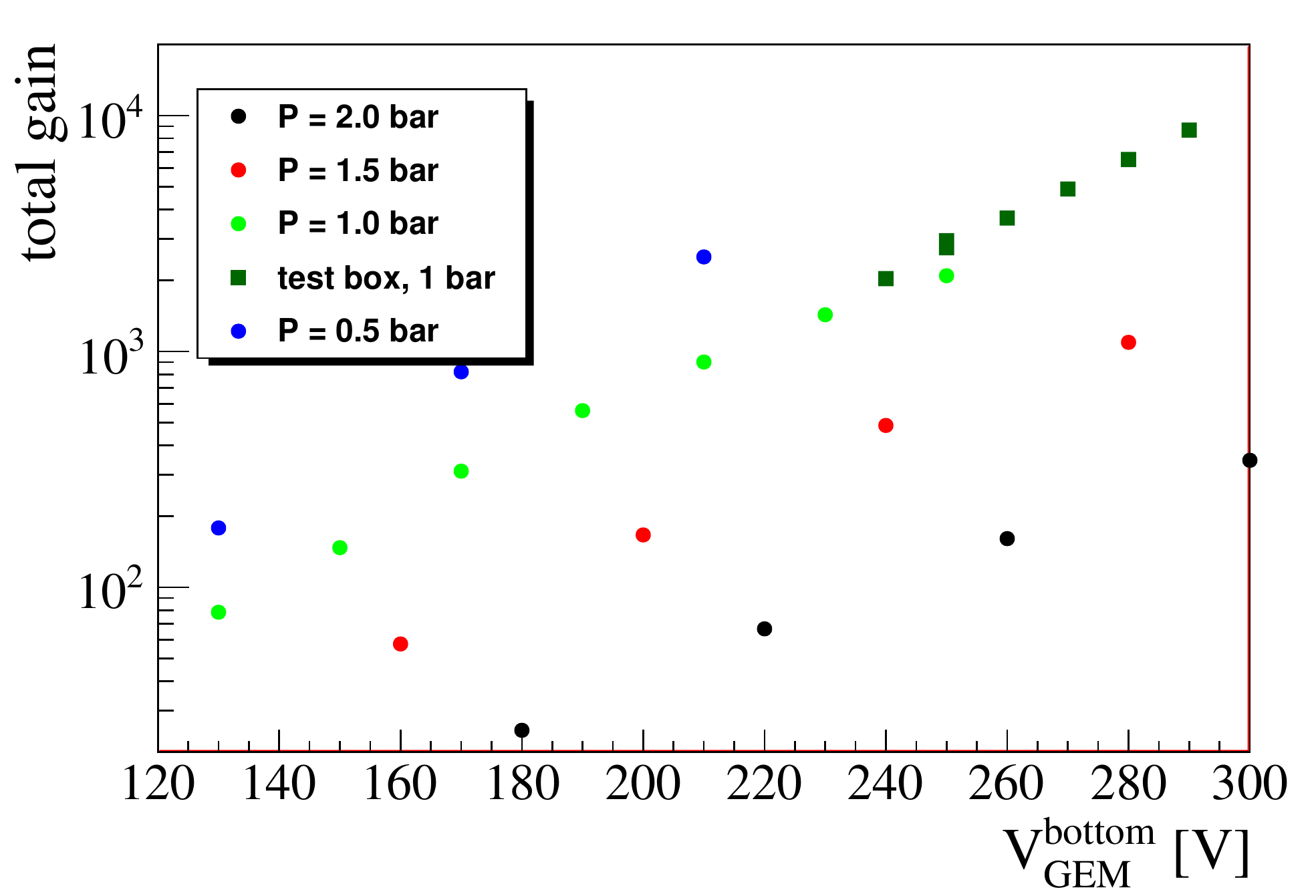}
 \includegraphics[width=0.32\linewidth]{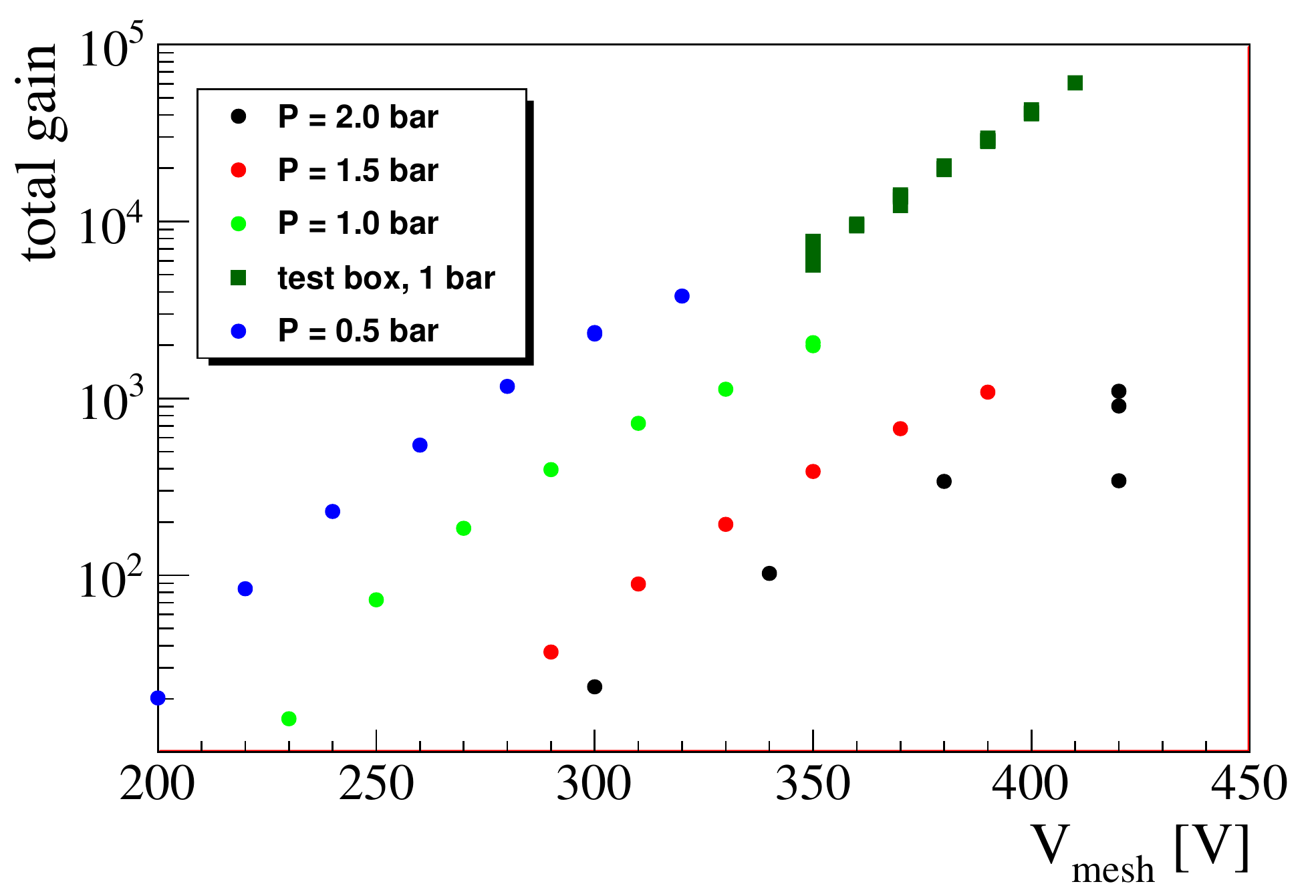} 
\caption{Gain in the HARPO TPC evaluated from traversing cosmic ray tracks\cite{PhGros:Feb2014:RD51}. 
}
\label{fig:GainCosmics}
\end{center}
\end{figure}

The $\micromegas + 2 ~ \GEM$ was commissioned into the TPC and the
full detector was tested with cosmic rays, using the same gas mixture.
Measurements of the energy deposit of tracks traversing the full
length of the TPC enable to evaluate the effective gain of the
amplification system.
Fig.~\ref{fig:GainCosmics} shows the dependence of the gain
with the voltage applied on the GEM and micromegas, for several gas
pressures.
The expected exponential behavior is confirmed.

\begin{figure} [Thb]
\begin{center}
 \setlength{\unitlength}{\textwidth}
 \begin{picture}(1,0.94)(0,0)



 \put(0,0.67){
 \includegraphics[width=0.49\linewidth]{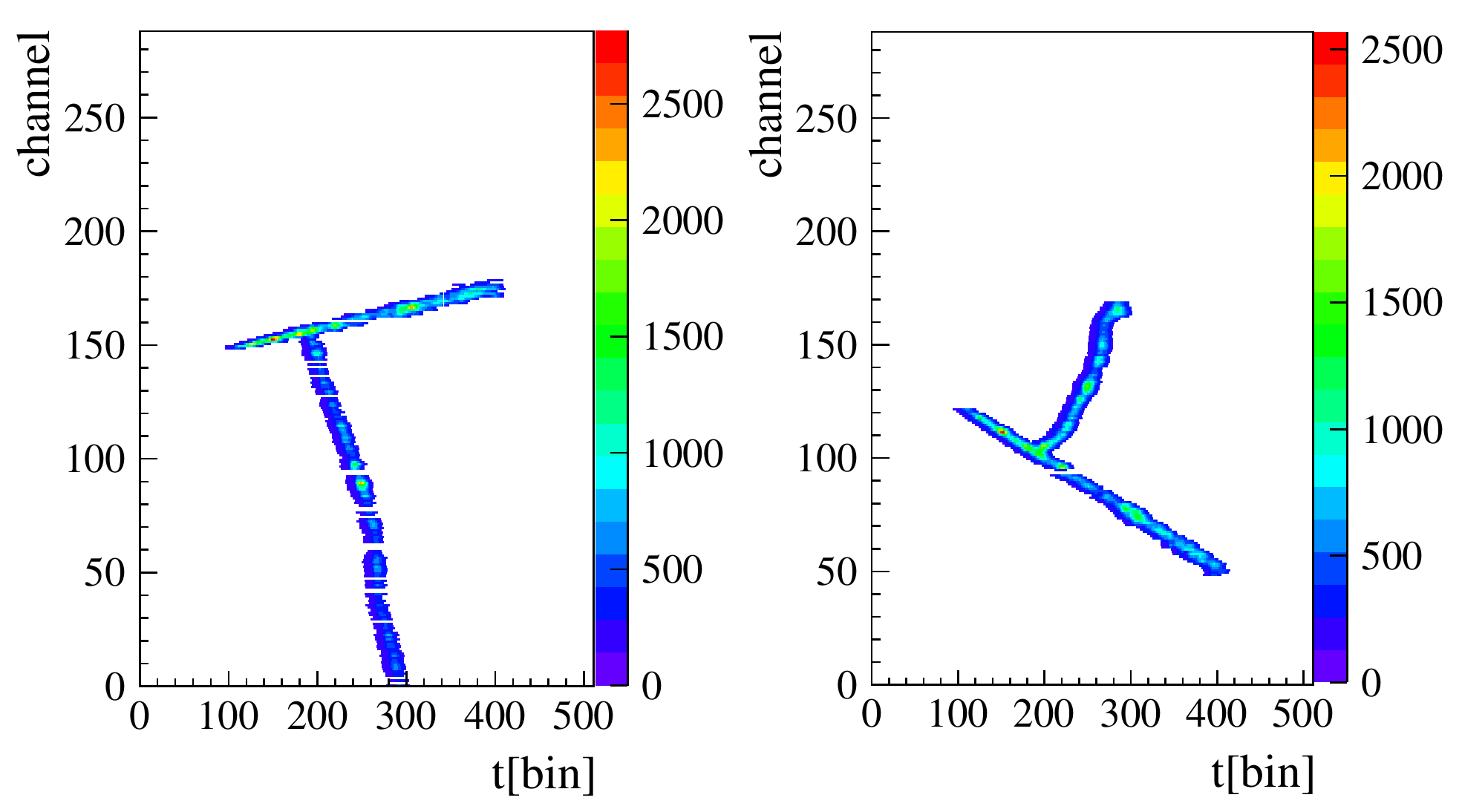}
\hfill
 \includegraphics[width=0.49\linewidth]{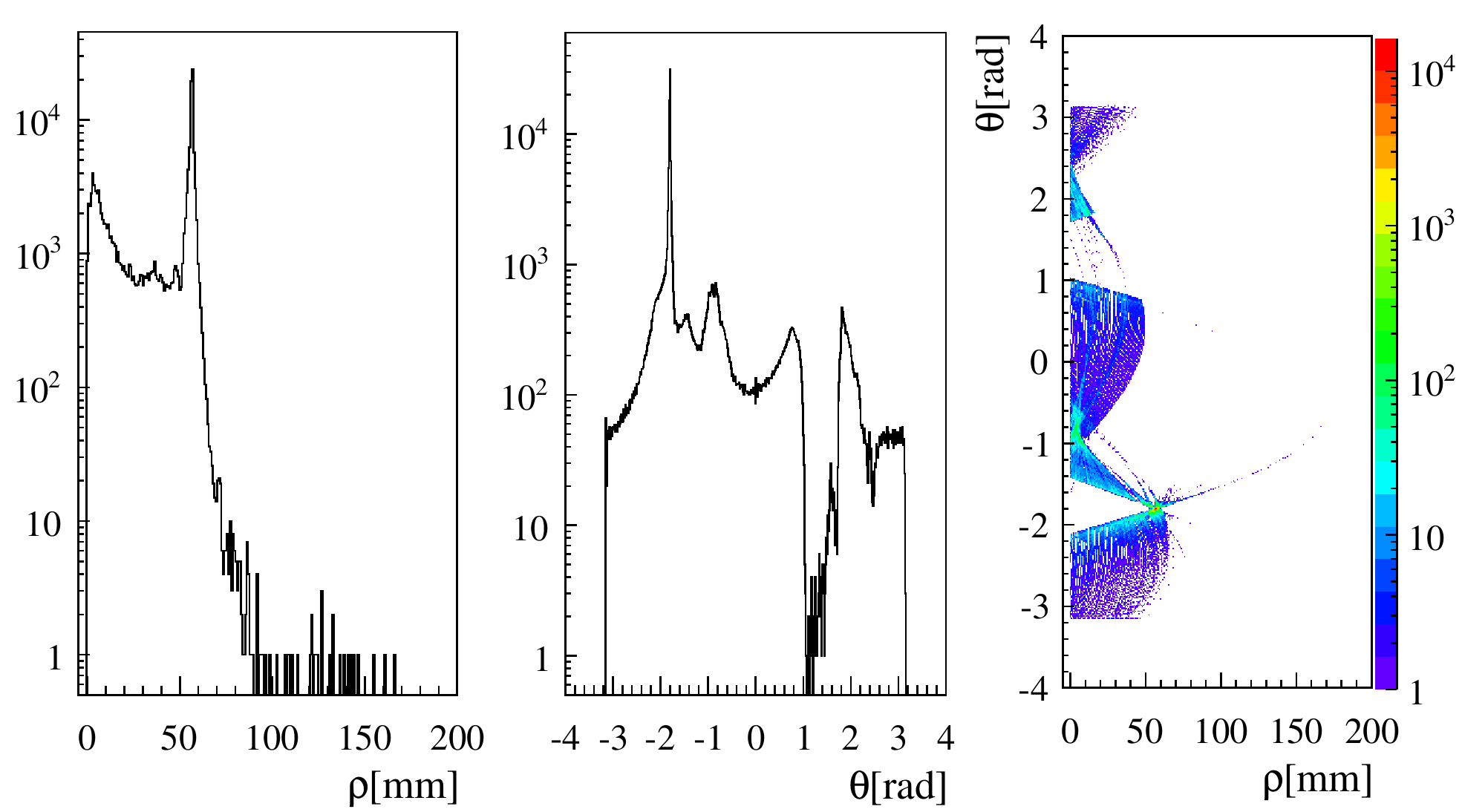}
}
 \put(0.17,0.89){\boldmath $x$}
 \put(0.41,0.89){\boldmath $y$}
 \put(0.24,0.94){\large \bf (a)}

 \put(0.63,0.89){\boldmath $y$}
 \put(0.79,0.89){\boldmath $y$}
 \put(0.94,0.89){\boldmath $y$}
 \put(0.73,0.94){\large \bf (b)}

 \put(0,0.39){
 \includegraphics[width=0.49\linewidth]{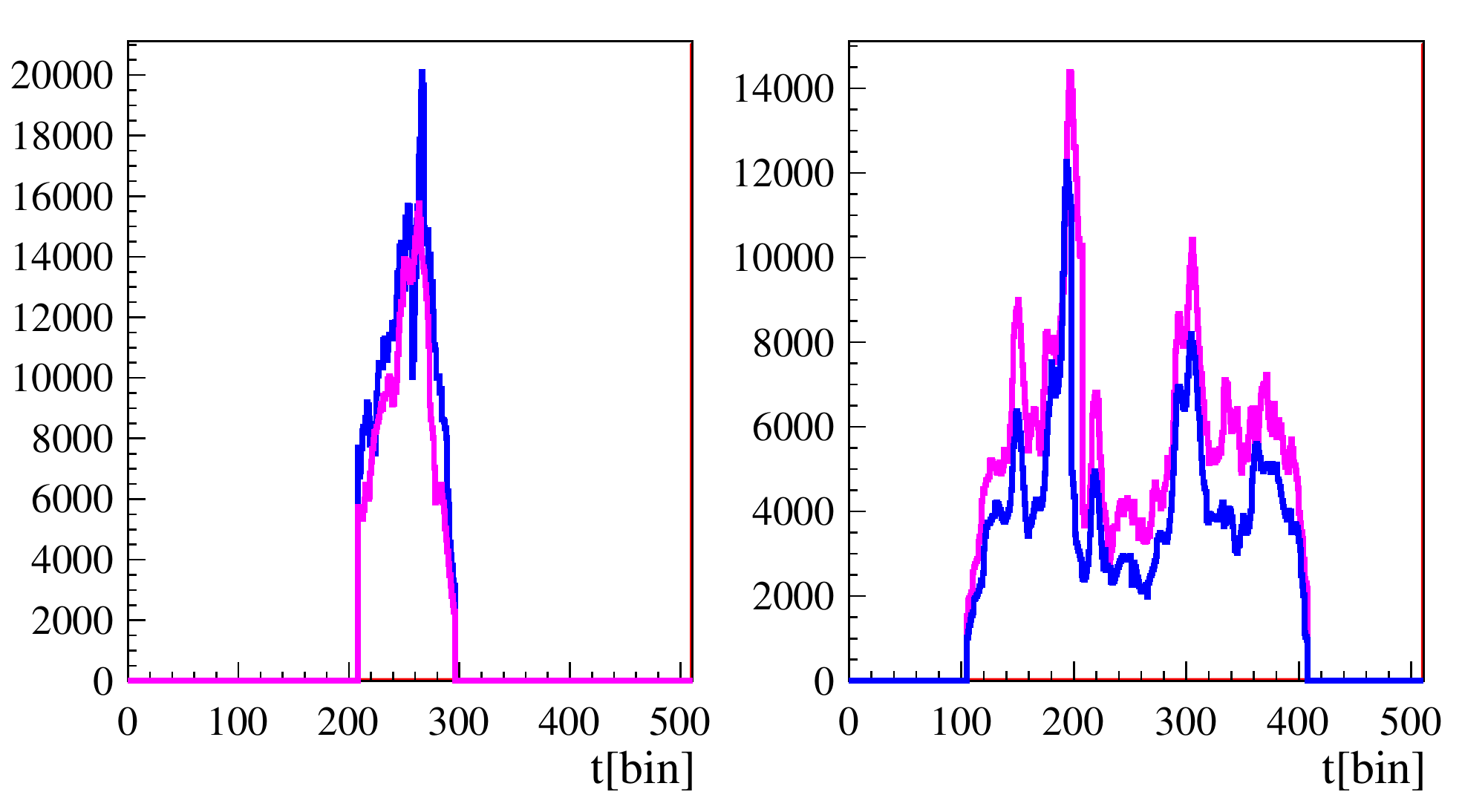}
\hfill
 \includegraphics[width=0.49\linewidth]{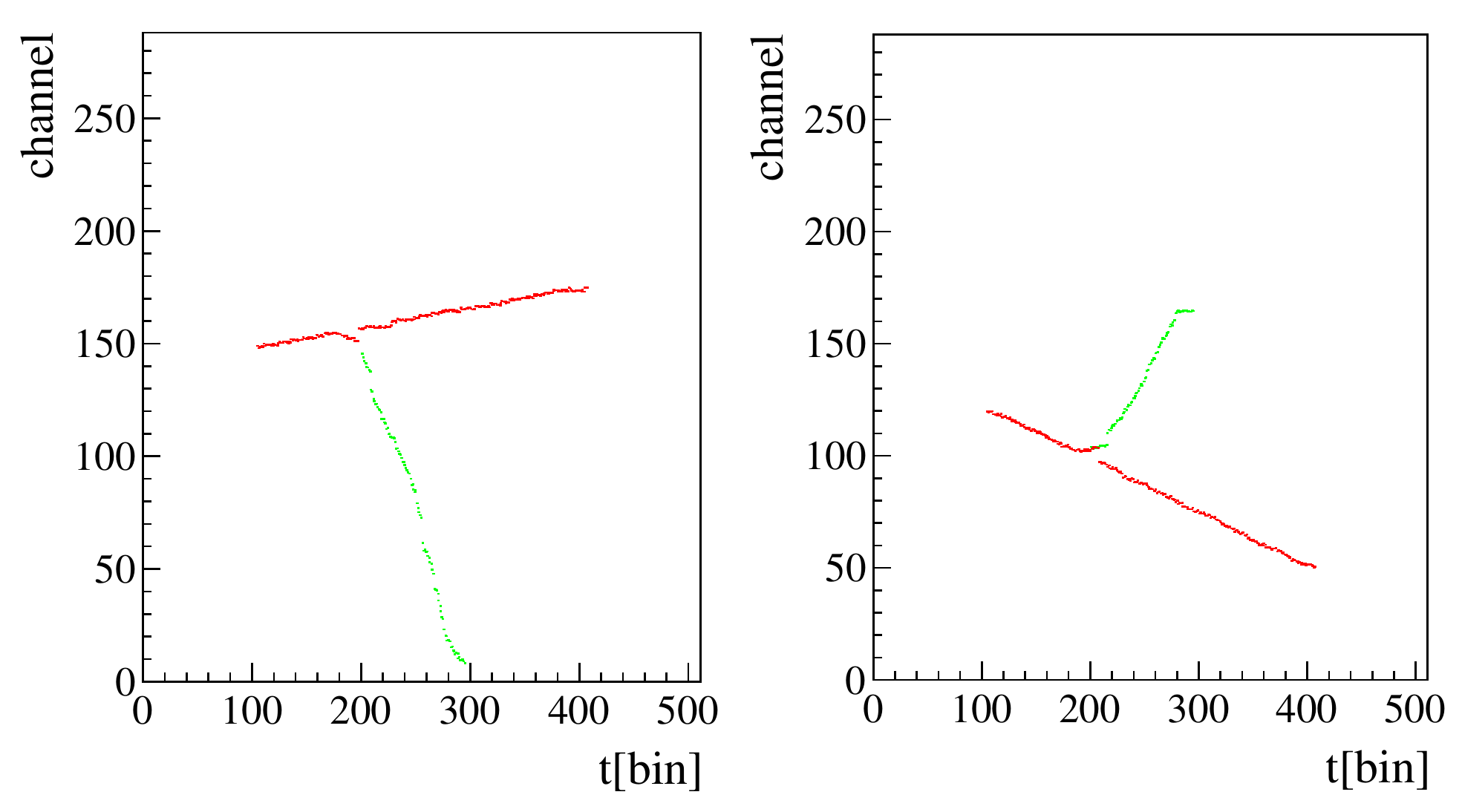}
}
\put (0.06,0.59) {\bf track 1}
\put (0.06,0.61) {\boldmath {\color{blue}$x$}, {\color{magenta}$y$}}

\put (0.31,0.59) {\bf track 2}
\put (0.31,0.61) {\boldmath {\color{blue}$x$}, {\color{magenta}$y$}}

\put (0.58,0.63) {\boldmath $x$}
\put (0.62,0.63) {\bf {\color{green}track 1}}
\put (0.62,0.60) {\bf {\color{red}track 2}}

\put (0.82,0.63) {\boldmath $y$}
\put (0.86,0.63) {\bf {\color{green}track 1}}
\put (0.86,0.60) {\bf {\color{red}track 2}}

 \put(0.24,0.66){\large \bf (c)}
 \put(0.73,0.66){\large \bf (d)}

 \put(0,-0.015){\centerline{
 \includegraphics[width=0.38\linewidth]{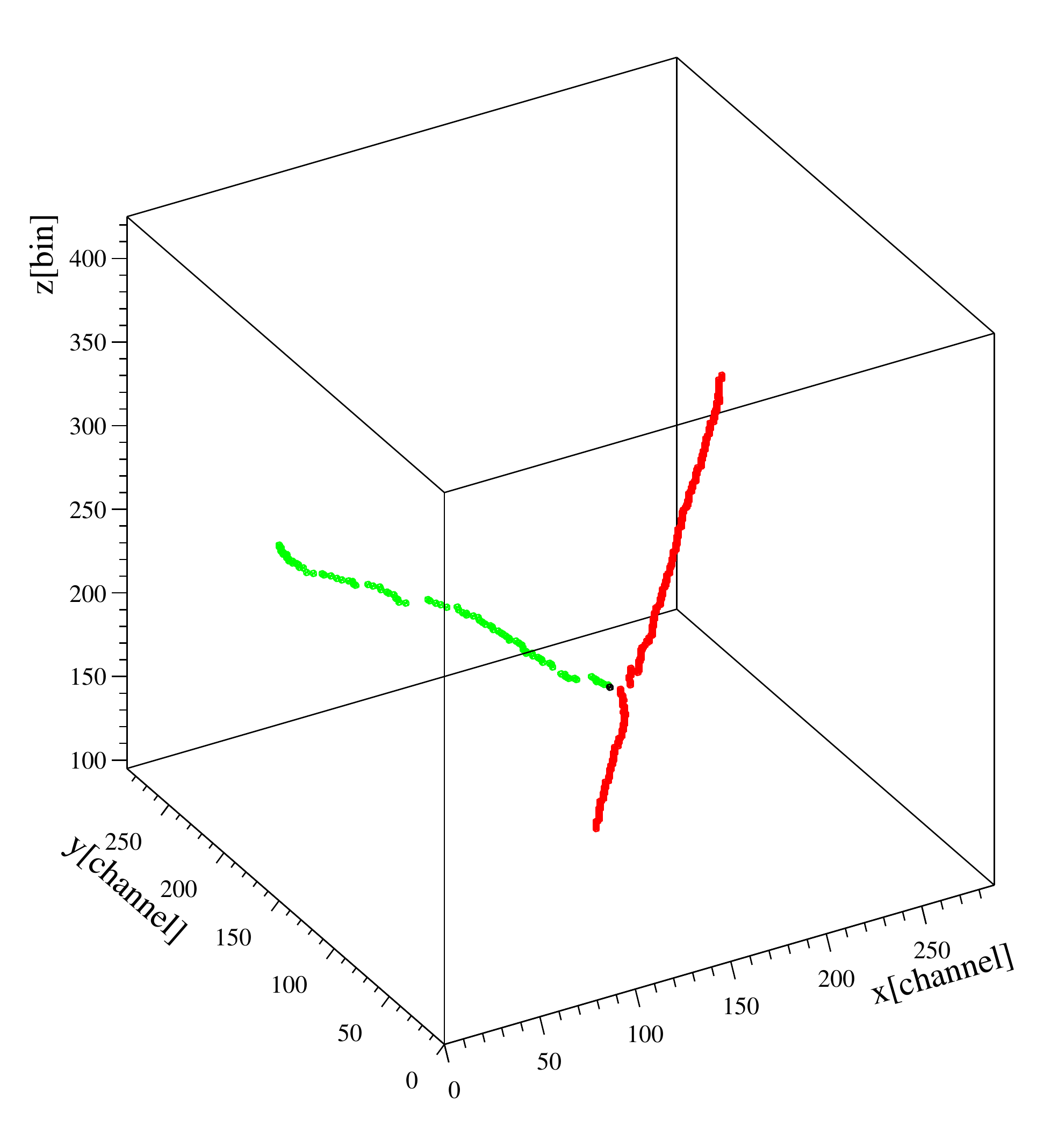}
 }}
 \put(0.57,0.2){\large \bf (e)}
 \end{picture}
\caption{Display of an event in the $\micromegas + 2 ~ \GEM$ detector
 configuration with a 2 bar (Ar:95 Isobutane 5 \%) gas mixture.
The TPC was positioned ``vertically'', with cosmic rays entering it
from above through the amplification system.
A cosmic ray crosses through the full TPC thickness (30cm), leaving a
lower energy electron ($\delta$ ray).
The trigger time for this run was $t_0 \approx 100 \bin$.
{\bf (a)} The two event ``maps'' ($x,t$) and ($y,t$) after pedestal
subtraction and thresholding (a thin space without any signal is
visible due to the presence of a spacer).
{\bf (b)} The distributions of the $\rho, \theta$ variables that
parametrize a track in the combinatorial Hough track pattern
recognition\cite{Bernard:2012jy,PhDShaobo} and their scatter plot.
{\bf (c)} The time spectra for the two tracks, for direction $x$ and
$y$, which are used to perform $x, y$ track matching.
{\bf (d)} The maps of the clusters associated with the two matched tracks.
{\bf (e)} A 3D view of the reconstructed event, with vertexing 
(the vertex is indicated by the black dot) \cite{PhDShaobo}.
\label{fig:Shaobo}
}
\end{center}
\end{figure}
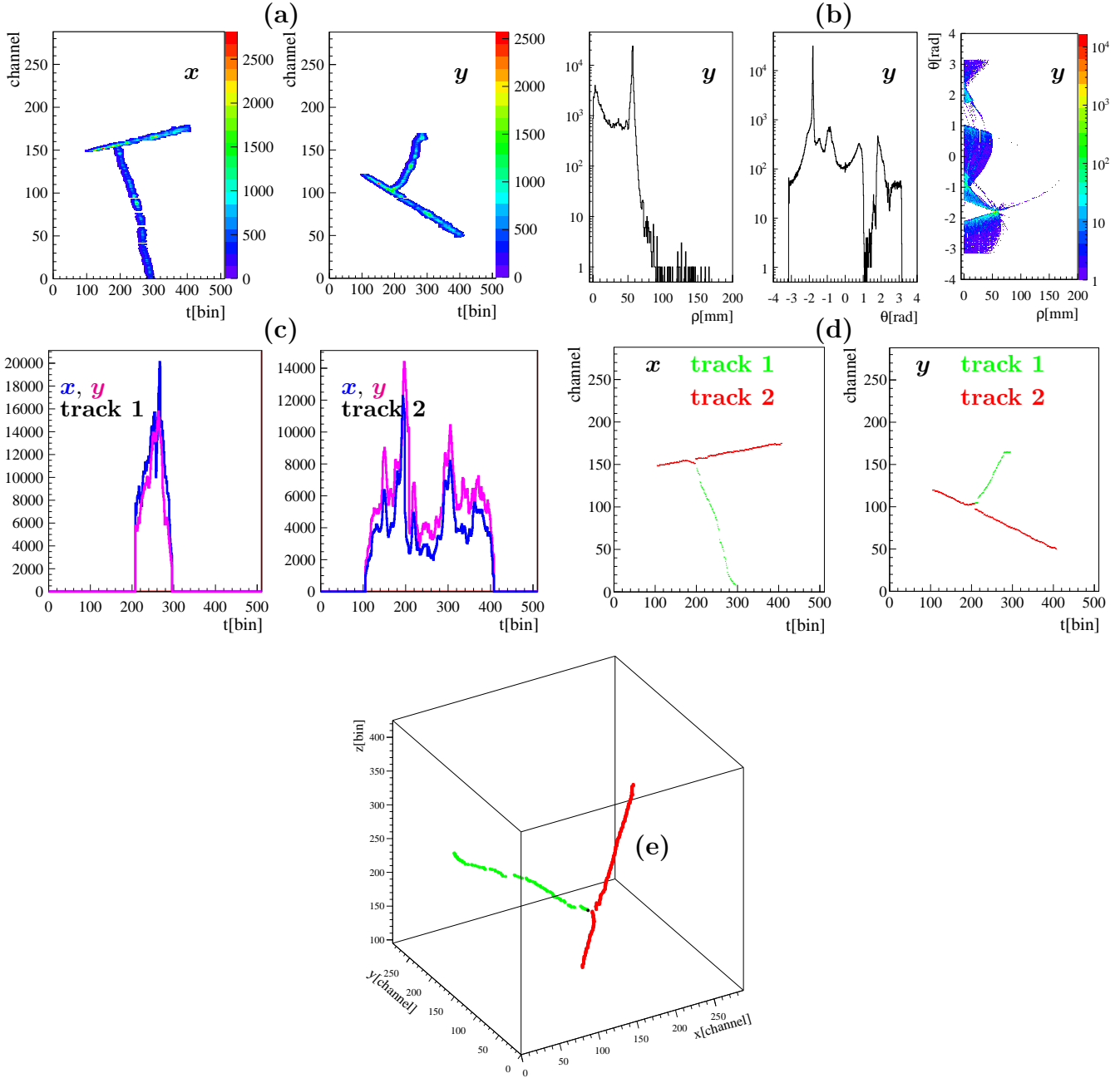

The test-box measurements and the TPC tests were performed with the same 
$\micromegas$ and $\GEM$s but with 
two
completely different electronics systems, that were calibrated separately.
It is therefore interesting to note that the variations of the signal
with $V_{\GEM}$ (Fig.~\ref{fig:GainCosmics} left and center) at 1 bar
show an agreement within $\approx 10\%$ between them.
At the highest value of the total gain, some saturation effects are
clearly visible (Fig.~\ref{fig:GainCosmics} right); their correction
is still under study.

\begin{figure} [th]
\begin{center}
 \includegraphics[width=0.532\linewidth]{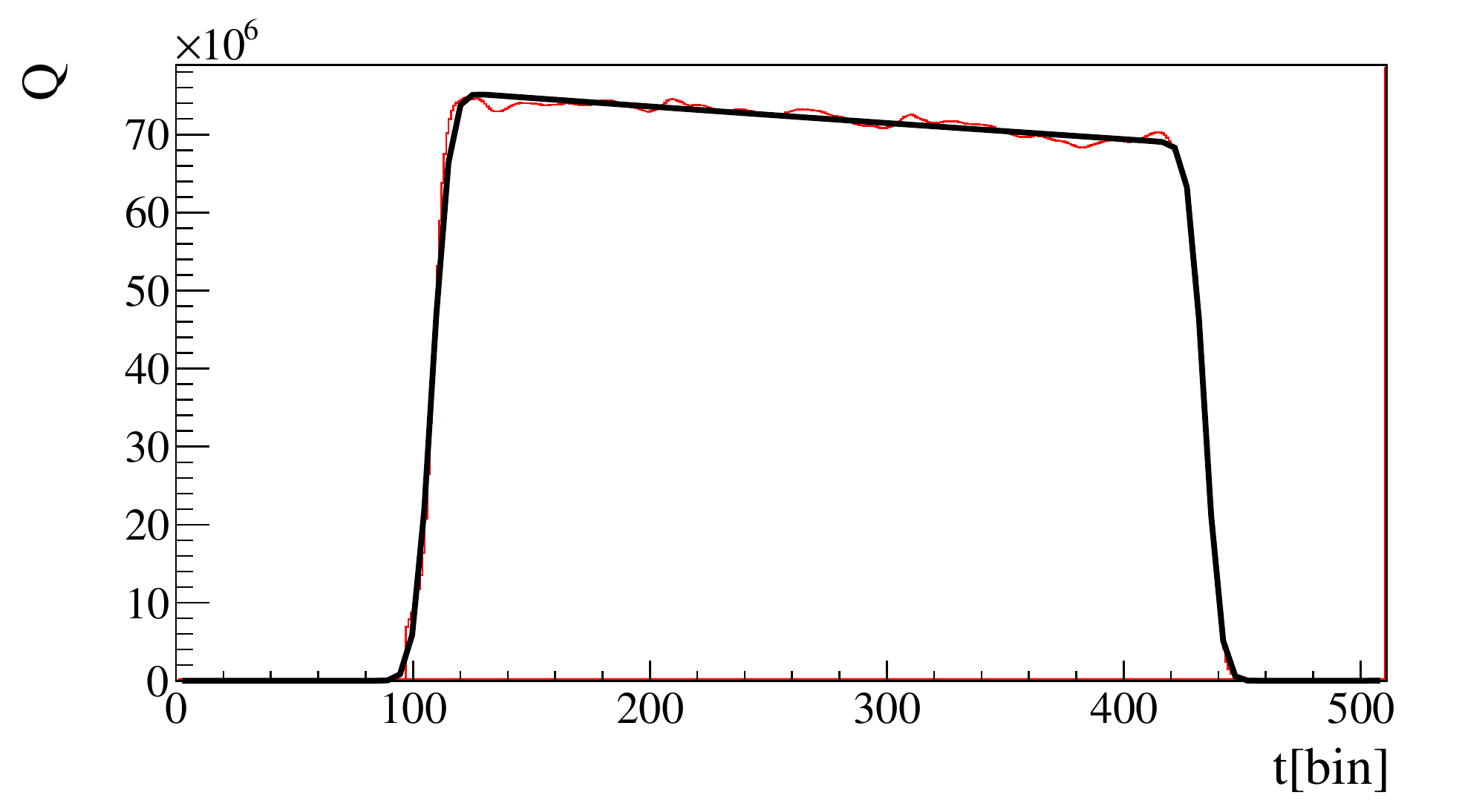}
\caption{Cumulative time spectrum (in ADC ``counts'') of
 traversing tracks of a run of 5000 events, in a (Ar:95 Isobutane:5
 \%) gas mixture at a pressure of 1.52 bar and with a sampling of 30
 ns and a shaping of 100 ns (Ref. \citenum{PhDShaobo}).}
\label{fig:TPC:calibration}
\end{center}
\end{figure}

\begin{figure} [th]
 \setlength{\unitlength}{\textwidth}
 \begin{picture}(1,0.65)(0,0)\thicklines
 \put(0,0.){
 \includegraphics[width=0.567\linewidth]{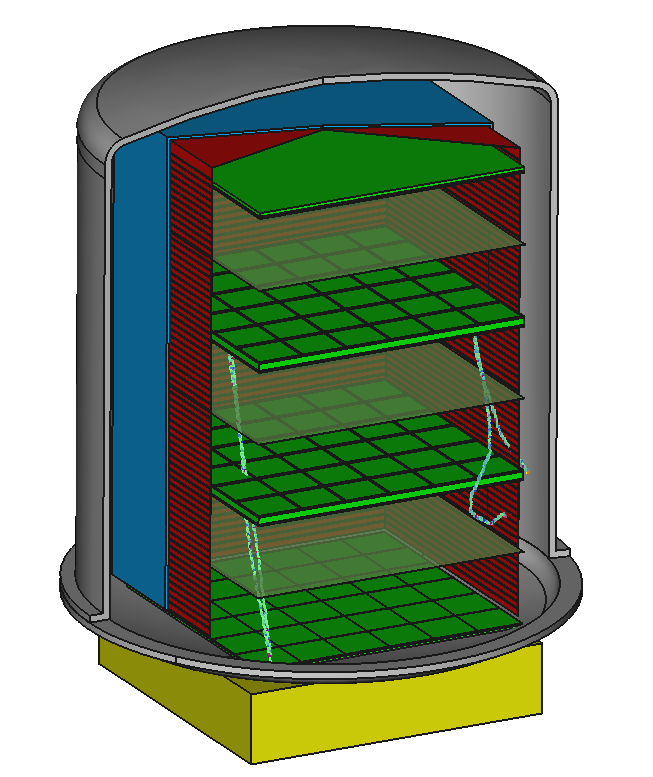}
}
 \put(0.4,0.08) {\line(6,1){0.25}}
 \put(0.67,0.12) {Service module}

 \put(0.43,0.18) {\line(6,1){0.23}}
 \put(0.67,0.22) {TPC field cage}

 \put(0.45,0.329) {\line(6,1){0.19}}
 \put(0.65,0.36) {Cathode}

 \put(0.45,0.395) {\line(6,1){0.13}}
 \put(0.6,0.42) {Anode, amplification \& FE electronics}

 \put(0.48,0.5) {\line(6,1){0.15}}
 \put(0.64,0.52) {Pressure vessel}

 \put(0.41,0.566) {\line(6,1){0.22}}
 \put(0.64,0.6) {Proton veto}
 \end{picture}
 \caption{Exploded schematic view of a flight telescope consisting of
  3 layers, each layer of 2 back-to-back modules, each module a
  $(2~\meter)^2 \times 0.5~\meter$ TPC with an endplate segmented
  into $(33~\centi\meter)^2$ micromegas and charge collection blocks.
Conversions of a 100 MeV (left) and of a 10 MeV (right) photon in the
TPC gas.
 \label{fig:schema:flight:detector}
}
 \end{figure}

Event reconstruction is documented in Fig. \ref{fig:Shaobo}.
The calibration of the TPC, i.e. the measurement of the trigger time
$t_0$ and of the drift velocity $v_{\drift}$ that take part in the basic
relation of the TPC mechanism, $z = v_{\drift} (t - t_0)$, is performed
easily with ``through'' tracks, that is with cosmic rays that cross the full $z$
thickness of 30 cm (Fig. \ref{fig:TPC:calibration}).
The low electron absorption rate of $10~\milli\second^{-1}$ along the
track is visible.
The combination of the values of the drift velocity and of the
digitization sampling rate, and the trigger delay are set so that the
``physical'' signal from the TPC extend from time bin 100 to bin
400, so that both in the transverse ($x, y$) and in the longitudinal
($z$) directions, the TPC is sampled with the same pitch of 1 mm.

The upgrade of the electronics includes the use of the recently
developed Feminos\cite{Feminos} back-end cards and of the PMm2 trigger
card \cite{Genolini:2008uc} based of the PARISROC
chip\cite{ConfortiDiLorenzo:2012sm}.

\section{Perspectives}

\paragraph{Short term 2014.}

We are presently finalizing the demonstrator set-up, with the goal of
characterizing it in a beam of polarized gamma rays at
NewSUBARU\cite{NewSUBARU} in the energy range 2 - 76 MeV in the fall
of this year 2014.

\paragraph{Longer term.}
We plan to develop the spatialization of the TPC
technology, with in particular studies of: 

\begin{itemize}
\item {\bf Triggerability}. A limitation of the present electronics
 based on the AFTER chip is the need of an external trigger -- at the
 moment provided by the combination of signals from scintillators.
We will replace these chips by the new generation,
self-triggerable member of the same family: the AGET chip \cite{AGET}
that has been tested successfully recently.
The full use of the real time information provided by the chip during
the drift of the electrons in the gas ($\approx 5~\micro\second$)
should allow us to build an efficient trigger much faster than the
digitization time ${\cal O} (\milli\second)$.

\item {\bf Clean technology}. In the present ``ground'' demonstrator,
 most of the detector components (TPC field cage, PCB, scintillators,
 wavelength shifter bars .. glue ..) are located inside the same sealed
 pressure vessel, without any gas purification.
As a consequence, after a month of running, the TPC showed a significant
absorption of the drifting electrons along their path through the
full length.
The use of a TPC in space will need to maintain the gas to a high
purity for several years.

Besides the standard use of purification of recirculated gas, the
content of the pressurized vessel could be split into two (series of)
separate volumes, a ``clean'' volume dedicated only to drift and
amplification, and a ``dirty'' part that contains all the rest,
strip PCB, front-end electronics (FE) .. 
The signal is read by its capacitive coupling through a
dielectric layer as for the Aleph calorimeter \cite{Decamp:1990jra} or
in the ``Piggyback'' technology that has been
validated recently\cite{Piggyback}.

\item {\bf Small gaps for high pressure operation.} 
Last but not least, the optimization of the sensitivity to faint
 sources and to polarization at a given limit volume encourages to
 use high-pressure detectors.
As the maximum of the micromegas gain is reached at a pressure that
increases when the micromegas gap size is decreased
\cite{Attie:2014pra}, we are developing small gap ``micro-bulk''
detectors \cite{Andriamonje:2010zz}.
\end{itemize}

\section*{Acknowledgments} 

Its our pleasure to acknowledge contributions from Leszek Ropelewski
{\it et al.} (RD51 lab at CERN) for the commissioning of the GEM,
Eric Delagnes and Denis Calvet (Irfu) for their help in using the T2K
electronics,
Eric Wanlin (IPNO) for his support on the PMm2 trigger card 
and Shuji Miyamoto (LASTI) for his continued support in the
preparation of the data taking at NewSUBARU.

This work is funded by the P2IO LabEx (ANR-10-LABX-0038) in the
framework ``Investissements d'Avenir'' (ANR-11-IDEX-0003-01) managed
by the French National Research Agency (ANR),
and directly by the ANR (ANR-13-BS05-0002).


\section{Notes added in proof}

During the conference I profited from many discussions relevant to
this paper:

\begin{itemize}
\item Keith Jahoda \cite{ref:GEMS:SPIE2014} pointed me to a study \cite{ref:GEMS:SPIE2013}  that
  demonstrated the feasability of a TPC detector with a lifetime on
  the 20 year scale for the GEMS project;

\item Eric Charles insisted on the need to improve the angular
  resolution in the $ < 1 \giga\electronvolt$ energy range in his
  post-Fermi ``lessons learned'' talk \cite{ref:Fermi:SPIE2014};

\item Meng Su presented the PANGU project \cite{ref:PANGU:SPIE2014}.
  One of its options is a stack of silicon detectors without any high
  $Z$ converter which, in contrast with previous works, deliberately
  attempts to perform polarimetry by the use of thin
  ($150~\micro\meter$) wafers.  Indeed, in the approximation of
  hyper-thin wafers spaced at a large enough distance, such a stack
  would become equivalent to a low density homogeneous detector, and
  the formalism of Refs.  \citenum{Bernard:2012uf,Bernard:2013jea}
  would apply.
\end{itemize}

\pagebreak

\tableofcontents

\end{document}